%% file: polarisation_paper_noB_sims.tex
\documentclass[aps,prd,amssymb,groupedaddress]{revtex4}
\usepackage{graphicx,amsmath}
\usepackage{epsf}
\usepackage{color}

\input Planck.tex

\usepackage{txfonts}

\usepackage{fixltx2e}
\usepackage{txfonts}
\usepackage{url}
\usepackage{natbib}
\bibpunct{[}{]}{;}{n}{}{,}  
\def\eqref#1{Eq.$\,$(\ref{#1})}

\newcommand{\eq}{\begin{align}}
\newcommand{\qe}{\end{align}}

\newcommand*\swallow[1]{}

\def\a{\alpha}

\def\D{\Delta}

\def\L{\Lambda}

\def\u{\mu}

\def\w{\omega}

\def\px{\approx}

\def\({\left(}
\def\){\right)}
\def\[{\left[}
\def\]{\right]}
\def\<{\left\langle}
\def\>{\right\rangle}

\def\un{{\bf \hat{n}}}

\def\bk{{\bf k}}

\def\curl{\mathcal}

\def\eq{\begin{align}}
\def\qe{\end{align}}

\def\and{\quad \mbox{and} \quad}

\def\bfnl{\kern2pt\overline{\kern-2ptf}_\textrm{NL}}

\def\lall{\ell_1\ell_2\ell_3}

\def\klist{k_1,k_2,k_3}

\def\Slll{S_{\ell_1\ell_2\ell_3}}

\def\barQ{\kern2pt\overline{\kern-2pt{Q}}}

\def\barR{\kern2pt\overline{\kern-2pt{R}}}

\definecolor{dblue}{rgb}{0.2,0,0.5}
\definecolor{dgreen}{rgb}{0.2,0.5,0}

\begin{document}

\title{Efficient optimal non-Gaussian CMB estimators with polarisation}

\author{J.R. Fergusson}
\affiliation
{Centre for Theoretical Cosmology,\\
Department of Applied Mathematics and Theoretical Physics,\\
University of Cambridge,
Wilberforce Road, Cambridge CB3 0WA, United Kingdom}

\begin{abstract}
In this paper we demonstrate an efficient method for including both CMB temperature and polarisation data in optimal non-Gaussian estimators. The method relies on orthogonalising the multipoles of the temperature and polarisation maps and results in a reduction by a factor of over 3 the terms required to calculate the estimator. The method is illustrated with the modal method applied to bispectrum estimation via the CMB with the trispectrum included as an appendix. However, the method is quite general and can be applied to any optimal bispectrum or trispectrum estimator including the KSW, binned and wavelet approaches. It would also be applicable to any situation where multiple data sets with known correlations are being considered.
\end{abstract}

\maketitle


\section{Introduction}
The paradigm of slow-roll single field inflation is now strongly favoured due to the recent results obtained by Planck \cite{1303.5082} and BICEP \cite{1403.3985}. One of the most promising areas to search for deviations from this standard model is through non-Gaussianities of the primordial density perturbation.  If detected, the form the non-Gaussianity took would point to specific mechanisms at play during inflation (see reviews \cite{1001.4707,1002.1416,1003.6097,1006.0275}). The recent Planck papers contained the strongest constraints on non-Gaussianity that currently exist \cite{1303.5084}.  No deviations from a Gaussian spectrum were found except some weak hints for oscillatory-type models.  As the temperature data is almost cosmic variance limited, for these constraints to be improved we need additional data sets.  In the future large scale structure may provide stronger constraints (see, for example, \cite{1206.1225}) but various theoretical and observational challenges remain before LSS becomes competitive with the CMB.  The easiest additional data set to include is the polarisation of the CMB. This has been measured by the current Planck satellite and will likely be part of the next release.  Hence the question of how to include polarisation data into current methods in an efficient manner is a pressing question.  Here we present an approach which greatly simplifies the complexity of the equations required for constraining non-Gaussianity via the bispectrum and trispectrum. The method is general to any optimal bi- or tri-spectrum estimator however we use the modal estimator as an illustrative example.

We begin by reviewing some basic equations for the CMB bispectrum which provide the starting point for our discussion. The primordial and CMB bispectrum are defined from the three-point correlators of the primordial density perturbation, $\Phi$, and the CMB multipoles, $a^X_{lm}$, respectively. For the CMB the superscript $X$ can be one of ($T$,$E$) denoting whether the multipoles were derived from temperature or E-mode polarisation CMB maps. Statistical isotropy demands that the primordial bispectrum has no angular dependence which translates to the CMB bispectrum having no $m$ dependence. Momentum conservation demands the three $\bk$ vectors form a triangle and this is enforced via a delta function. This leads to the following expressions:
\begin{align}
\< \Phi(\bk_1) \Phi(\bk_2) \Phi(\bk_3)\> &= \(2\pi\)^3\delta(\bk_1+\bk_2+\bk_3) B(k_1,k_2,k_3)\,,\\
\< a^{X_1}_{\ell_1 m_1} a^{X_2}_{\ell_2 m_2} a^{X_3}_{\ell_3 m_3} \> &= \curl{G}^{\ell_1 \ell_2 \ell_3}_{m_1 m_2 m_3} b^{X_1X_2X_3}_{\ell_1 \ell_2 \ell_3}\,,
\end{align}
and $\curl{G}$ is the Gaunt integral, which is the projection of the angular part of the primordial delta function, and is defined as follows
\begin{align}
\nonumber\curl{G}^{\ell_1 \ell_2 \ell_3}_{m_1 m_2 m_3} &= \int d\Omega_{\un} Y_{\ell_1 m_1 }(\un) Y_{\ell_2 m_2 }(\un) Y_{\ell_3 m_3 }(\un) = \(\begin{array}{ccc}\ell_1 & \ell_2 & \ell_3 \\ m_1 & m_2 & m_3  \end{array}\) h_{\ell_1 \ell_2 \ell_3}\,, \\
h_{\ell_1 \ell_2 \ell_3} &= \sqrt{\frac{(2\ell_1+1)(2\ell_2+1)(2\ell_3+1)}{4\pi}}\(\begin{array}{ccc}\ell_1 & \ell_2 & \ell_3 \\ 0 & 0 & 0  \end{array}\)\,.
\end{align}
If we wish to constrain the amplitude of the bispectrum from the CMB we need to construct an estimator. The simplest form was written down and shown to be optimal in \cite{0503375}.  A linear term was added in \cite{0509029} to restore optimality in the presence on anisotropic noise and sky cuts. This estimator was then extended to include polarisation in \cite{0701921,0711.4933}. In its most general form the estimator is
\begin{align}\label{eq:estimator}
\nonumber \curl{E} &= \frac{1}{N}\sum_{X^{\phantom{'}}_i X'_i}\sum_{\ell^{\phantom{'}}_i \ell'_i m^{\phantom{'}}_i m'_i}  \curl{G}^{\ell^{\phantom{'}}_1 \ell^{\phantom{'}}_2 \ell^{\phantom{'}}_3}_{m^{\phantom{'}}_1 m^{\phantom{'}}_2 m^{\phantom{'}}_3} b^{X^{\phantom{'}}_1X^{\phantom{'}}_2X^{\phantom{'}}_3}_{\ell^{\phantom{'}}_1 \ell^{\phantom{'}}_2 \ell^{\phantom{'}}_3} (C^{-1})^{X^{\phantom{'}}_1X'_1}_{\ell^{\phantom{'}}_1\ell'_1m^{\phantom{'}}_1m'_1} (C^{-1})^{X^{\phantom{'}}_2X'_2}_{\ell^{\phantom{'}}_2\ell'_2m^{\phantom{'}}_2m'_2} (C^{-1})^{X^{\phantom{'}}_3X'_3}_{\ell^{\phantom{'}}_3 \ell'_3 m^{\phantom{'}}_3 m'_3}\\
&\times \(a^{X'_1}_{\ell'_1 m'_1} a^{X'_2}_{\ell'_2 m'_2} a^{X'_3}_{\ell'_3 m'_3} - \<a^{X'_1}_{\ell'_1 m'_1} a^{X'_2}_{\ell'_2 m'_2}\> a^{X'_3}_{\ell'_3 m'_3} - \<a^{X'_2}_{\ell'_2 m'_2} a^{X'_3}_{\ell'_3 m'_3}\> a^{X'_1}_{\ell'_1 m'_1} - \<a^{X'_1}_{\ell'_1 m'_1} a^{X'_3}_{\ell'_3 m'_3}\> a^{X'_2}_{\ell'_2 m'_2}\)\,,
\end{align}
where the normalisation, $N$, is defined by
\begin{align}\label{eq:norm}
N \equiv \sum_{X^{\phantom{'}}_i X'_i}\sum_{\ell^{\phantom{'}}_i\ell'_i}\curl{G}^{\ell^{\phantom{'}}_1 \ell^{\phantom{'}}_2 \ell^{\phantom{'}}_3}_{m^{\phantom{'}}_1 m^{\phantom{'}}_2 m^{\phantom{'}}_3} b^{X^{\phantom{'}}_1 X^{\phantom{'}}_2 X^{\phantom{'}}_3}_{\ell^{\phantom{'}}_1 \ell^{\phantom{'}}_2 \ell^{\phantom{'}}_3} (C^{-1})^{X^{\phantom{'}}_1X'_1}_{\ell^{\phantom{'}}_1\ell'_1m^{\phantom{'}}_1m'_1} (C^{-1})^{X^{\phantom{'}}_2X'_2}_{\ell^{\phantom{'}}_2\ell'_2m^{\phantom{'}}_2m'_2} (C^{-1})^{X^{\phantom{'}}_3X'_3}_{\ell^{\phantom{'}}_3\ell'_3m^{\phantom{'}}_3m'_3} \curl{G}^{\ell'_1 \ell'_2 \ell'_3}_{m'_1 m'_2 m'_3} b^{X'_1 X'_2 X'_3}_{\ell'_1 \ell'_2 \ell'_3}\,,
\end{align}
and $(C^{-1})^{XX'}_{\ell \ell' m m'}$ is the $XX'$ element of the inverse of the covariance matrix for the $a^X_{lm}$
\begin{align}\label{eq:covariance}
\(\begin{array}{cc}
C^{TT}_{\ell \ell' m m'} & C^{TE}_{\ell \ell' m m'} \\
C^{TE}_{\ell \ell' m m'} & C^{EE}_{\ell \ell' m m'} \,.
\end{array}\)^{-1}
\end{align}
The normalisation is related to the Fisher matrix by $N = 6F$. We can relate the primordial and CMB bispectra by a convolution
\begin{align}\label{eq:projection}
b^{X_1X_2X_3}_{\ell_1 \ell_2 \ell_3} = \(\frac{2}{\pi}\)^3\int_{\curl{V}_k} \(k_1 k_2 k_3\)^2 B(\klist) \D^{X_1 X_2 X_3}_{\ell_1 \ell_2 \ell_3}(k_1,k_2,k_3)  d\curl{V}_k\,,
\end{align}
where $d\curl{V}_k$ is the region of k space allowed by the triangle condition and we have defined the bispectrum transfer function $\D^{X_1 X_2 X_3}_{\ell_1 \ell_2 \ell_3}$ to be
\begin{align}
\D^{X_1 X_2 X_3}_{\ell_1 \ell_2 \ell_3}(k_1,k_2,k_3) \equiv \D^{X_1}_{\ell_1}(k_1) \D^{X_2}_{\ell_2}(k_2) \D^{X_3}_{\ell_3}(k_3) \int x^2 dx j_{\ell_1}(xk_1) j_{\ell_2}(xk_2) j_{\ell_3}(xk_3)\,.
\end{align}
where the $\D_l$ are the radiation transfer functions as produced by CMB anisotropy codes such as CAMB \cite{0205436}. The integral over the spherical Bessel functions $j_l$ is a geometric factor from the projection of the radial part of the primordial delta function. The above equations, while seemingly simple, are impossible to evaluate in general (excepting very small $l_{max}$). The convolution to calculate a single $\ell$-triple of $b_{\ell_1 \ell_2 \ell_3}$ requires a 4D integration and there are in general $\ell^3/2$ triples to calculate.  Also the estimator requires a sum over $\ell^{11}$ terms, which reduces to $\ell^5$ if we assume the covariance matrices are diagonal.  This is beyond current computational resources. Fortunately the equations simplify greatly if we consider primordial bispectra which are separable. This fact was exploited, in multiple ways relating to different choices of separable functions to filter the data with, in the recent Planck experiment to obtain constraints on a wide variety of primordial \cite{1303.5084} and late time models \cite{1303.5079, 1303.5085}. The methods fall into 4 main categories, KSW-type \cite{0302223, 0305189}, Binned \cite{0911.1642}, Modal \cite{0612713,0912.5516} and Wavelet \cite{9808.3987,0111284,0211399} approaches. Each approach has its own particular advantages and disadvantages which are briefly discussed later on. Non-Gaussianity can also be constrained with other estimators, like Minkowski functionals \cite{0302223,0401276}, but these are suboptimal and so the method we describe in this paper is not applicable to them.

In the following section we will review the temperature-only modal approach as used by Planck before demonstraiting a novel method for including polarisation in the subsequent section which comprises the focus of this paper. The choice of the modal approach is for illustration only and the method could be applied equally to any of the other methods.  This method is also applicable in the case of trispectrum estimation and the extension to it is included as an appendix.

\section{Modal methods for CMB temperature data}

The modal approach to non-Gaussian estimation relies on using two separable orthonormal bases, one at primordial time and one at the level of the data \cite{0912.5516}. This has been demonstrated to provide a numerically efficient method to constrain large numbers of primordial models simultaneously \cite{1006.1642}. Primordial models are first represented as a series of mode coefficients in the primordial basis. These can then be projected via matrix transform, which includes all necessary details of radiation transfer functions, to new mode coefficients in the CMB basis. The data is similarly compressed into mode coefficients in the CMB basis. The estimator then simply becomes the suitably normalised dot product of the model and data coefficients.

We must first define shape functions which are the quantities which we want to decompose. We choose them to closely match the signal-to-noise ratio for a given experiment. As the primordial bispectrum scales as $k^{-6}$ equation, \eqref{eq:projection}, provides a natural definition for the primordial shape function
\begin{align}
S(\klist) \equiv \(k_1 k_2 k_3\)^2 B(\klist)\,,
\end{align}
so the reduced CMB bispectrum is simply the projection of the primordial shape function by the bispectrum transfer function
\begin{align}\label{eq:shapeprojection}
b^{TTT}_{\ell_1 \ell_2 \ell_3} = \int_{\curl{V}_k} S(\klist) \D^{TTT}_{\ell_1 \ell_2 \ell_3}(k_1,k_2,k_3) d\curl{V}_k\,.
\end{align}
We will consider temperature data and work in the diagonal covariance approximation so $C^{XX'}_{\ell \ell' m m'} \px C^{XX'}_{\ell} \delta_{\ell \ell'} \delta_{m -m'}$. The normalisation of the estimator, \eqref{eq:norm}, points to a sensible choice for the CMB shape function
\begin{align}
\sqrt{\frac{h^2_{\ell_1 \ell_2 \ell_3}}{C^{TT}_{\ell_1} C^{TT}_{\ell_2} C^{TT}_{\ell_3}}}\, b^{TTT}_{\ell_1 \ell_2 \ell_3}\,.
\end{align}
However $h$ contains Wigner 3j symbols making it difficult to calculate. It is also non-separable which is important for reducing computational complexity as we will see later.  Instead we approximate it with $v_{\ell_1}v_{\ell_2}v_{\ell_3}$ where $v_{\ell} = (2\ell+1)^{1/6}$. This scales with $\ell$ in the same way as $h$ does providing a reasonable approximation. This choice defines our CMB shape function as
\begin{align}\label{eq:CMBshape}
S^{TTT}_{\ell_1 \ell_2 \ell_3} \equiv \sqrt{\frac{v^2_{\ell_1}v^2_{\ell_2}v^2_{\ell_3}}{C^{TT}_{\ell_1} C^{TT}_{\ell_2} C^{TT}_{\ell_3}}}\,b^{TTT}_{\ell_1 \ell_2 \ell_3}\,.
\end{align}
We now define the primordial and CMB inner products which we use to create our orthonormal basis as
\begin{align}
\<A,\, B\>_k &\equiv  \int_{\curl{V}_k} \, \w(k_1,k_2,k_3) A(k_1,k_2,k_3) \, B(k_1,k_2,k_3) d\curl{V}_k \,,\\
\<A,\, B\>_l &\equiv \sum_{\ell_i} \w_{\ell_1 \ell_2 \ell_3} \,A_{\ell_1 \ell_2 \ell_3} \, B_{\ell_1 \ell_2 \ell_3}\,,
\end{align}
with the corresponding weights given by
\begin{align}
\bar\w(k_1,k_2,k_3) = \frac{1}{k_1+k_2+k_3}\,,\qquad
\w_{\ell_1 \ell_2 \ell_3} = \left( \frac{h_{\ell_1 \ell_2 \ell_3}}{v_{\ell_1}v_{\ell_2}v_{\ell_3}}\right)^2\,.
\end{align}
The weight in $\ell$-space is such that the inner product of two CMB shape functions is proportional to the Fisher matrix. The $1/k$ scaling of the primordial weight is to reflect the $1/l$ scaling of the Fisher matrix which has been shown to be necessary to obtain accurate predictions for primordial models which are not scale invariant \cite{0812.3413}.

We require two sets of separable basis functions or eigenmodes, one defined at primordial times $\bar{Q}_n(k_1,k_2,k_3)$ and one at late times $Q_{n\,l_1 l_2 l_3}$. We employ these to expand the respective shape functions as
\begin{align}\label{eq:Sprimexp}
\bar{S}(\klist) &= \sum_n \bar{\alpha}_n\,\bar{Q}_n(\klist)\,,\\
\Slll &= \sum_n \alpha_n \,Q_{n\,l_1 l_2 l_3}\,,\label{eq:Scmbexp}
\end{align}
where we have denoted primordial quantities by placing a bar over them. The expansion coefficients $\bar\alpha _n$ and $\alpha_n$ are given by
\begin{align}\label{eq:Qcoeff}
\bar{\alpha}_n= \sum_p\bar{\gamma}^{-1}_{np}\<\bar{Q}_p,\, \bar{S}\>_k \,,\qquad \alpha_n = \sum_p\gamma^{-1}_{np} \<Q_p,\, S\>_\ell\,,
\end{align}
with the $\gamma$-matrices formed from the non-orthogonal $Q$ products:
\begin{align}
\bar{\gamma}_{np} = \<\bar{Q}_n,\, \bar{Q}_p\>_k\,,\quad \qquad \gamma_{np} = \<Q_n, \,Q_p\>_\ell \,.
\end{align}
For convenience, these modes can be orthonormalised to form new basis sets $\bar{R}_n(k_1,k_2,k_3)$ and $R_{n\,l_1 l_2 l_3}$, such that
\begin{align}\label{eq:Rmodes}
 \<\bar{R}_n,\, \bar{R}_p\>_k ={\delta}_{np}\,, \qquad  \<R_n,\, R_p\>_\ell = \delta_{np}\,.
\end{align}
This is done by taking the Cholesky decomposition of the associated $\gamma^{-1}$,
\begin{align}
\bar{\gamma}^{-1}_{np} = \sum_r \bar{\lambda}_{nr}\bar{\lambda}_{pr} \,,\qquad \gamma^{-1}_{np} = \sum_r \lambda_{nr} \lambda_{pr}\,,
\end{align}
where the $\lambda$ are lower triangular. This is equivalent to performing a modified Gram-Schmidt orthogonalisation but is more numerically stable and efficient. An alternative approach would be to use principal component analysis and decompose $\gamma$ into its eigenvalues and eigenvectors so, $\gamma = V D V^T$, where $V$ is the matrix of unit eigenvectors and $D$ is diagonal with the eigenvalues on the diagonal.  Then $\lambda$ is defined by $\lambda = V\sqrt{1/D}$ as $V^{-1} = V^T$. For unstable bases we can optionally truncate the calculation of $\lambda$ to only consider eigenvalues over a certain threshold so $\lambda$ becomes rectangular.

We can relate the non-orthogonal and orthonormal modes through
\begin{align}
\bar{R}_n = \sum_p \bar{\lambda}_{np} \bar{Q}_p \,,\qquad R_n = \sum_n \lambda_{np} Q_n \,.
\end{align}
Thus the model expansion coefficients $\alpha^R_n$ in the orthonormal basis are given in terms of the original expansions, \eqref{eq:Qcoeff}, by
\begin{align}\label{eq:Rcoeff}
\bar{\alpha}^R_n= \sum_p\bar{\lambda}^{-1}_{pn} \bar{\alpha}_p\,,\qquad \alpha^R_n = \sum_p\lambda^{-1}_{pn} \alpha_p\,.
\end{align}

The primordial basis $Q_n(\klist)$ can be projected via the bispectrum transfer function to the corresponding late-time functions $\widetilde Q_{n\,\lall}$, where we have used a tilde to denote that they are the result of projection,
\begin{align}\label{eq:projbasis}
\widetilde{Q}^{TTT}_{n\,l_1 l_2 l_3} = \sqrt{\frac{v^2_{\ell_1}v^2_{\ell_2}v^2_{\ell_3}}{C^{TT}_{\ell_1} C^{TT}_{\ell_2} C^{TT}_{\ell_3}}} \int_{\curl{V}_k}\bar{Q}_n(k_1,k_2,k_3)\, \Delta^{TTT}_{\ell_1 \ell_2 \ell_3}(k_1,k_2,k_3) d\curl{V}_k \,.
\end{align}
This choice is made so that if the primordial shape function can be described by
\begin{align}
\bar{S}(\klist) &= \sum_n \bar{\alpha}_n\,\bar{Q}_n(\klist)\,,
\end{align}
then the equivalent expression holds in $\ell$-space with the projected quantities,
\begin{align}
\Slll &= \sum_n \bar{\alpha}_n \,\widetilde{Q}_{n\,l_1 l_2 l_3}\,.
\end{align}

These projected early-time modes $\widetilde Q_n$ can be related to the late-time modes $Q_n$ through a transformation matrix $\Gamma_{np}$ defined as
\begin{align}
\Gamma_{np} = \sum_r\gamma^{-1}_{nr} \<Q_r, \,\widetilde{Q}_p\>_\ell\,.
\end{align}
Hence, having set up this modal machinery and calculated $\Gamma_{np}$,  we can now project the expansion coefficients $\alpha_n$ of an arbitrary primordial bispectrum, \eqref{eq:Sprimexp}, to obtain the corresponding CMB bispectrum, \eqref{eq:Scmbexp}, with expansion coefficients given by
\begin{align}\label{eq:earlylatealpha}
\alpha_n =\sum_p\Gamma_{np}\,\bar\alpha_p\,.
\end{align}
In principle, the inverse of $\Gamma_{np}$ can be used to gain insight about the primordial bispectrum from the measured CMB bispectrum.

If we replace the CMB shape function with our expansion into modes in the expression for the temperature only estimator we obtain.
\begin{align}
\curl{E} &= \frac{1}{N} \sum_{\ell m}  \frac{\curl{G}^{\ell_1 \ell_2 \ell_3}_{m_1 m_2 m_3}}{v_{\ell_1}v_{\ell_2}v_{\ell_3}}  \frac{ \sum_n \alpha^{TTT}_n Q^{TTT}_{n \ell_1 \ell_2 \ell_3}}{\sqrt{C^{TT}_{\ell_1} C^{TT}_{\ell_2} C^{TT}_{\ell_3}}}\bigg(a^{T}_{\ell_1 m_1} a^{T}_{\ell_2 m_2} a^{T}_{\ell_3 m_3} - \<a^{T}_{\ell_1 m_1} a^{T}_{\ell_2 m_2}\> a^{T}_{\ell_3 m_3} - \<a^{T}_{\ell_2 m_2} a^{T}_{\ell_3 m_3}\> a^{T}_{\ell_1 m_1} - \<a^{T}_{\ell_1 m_1} a^{T}_{\ell_3 m_3}\> a^{T}_{\ell_2 m_2}\bigg)\,.
\end{align}
Now if we define the CMB data shape function as
\begin{align}
S^{TTT}_{D\ell_1 \ell_2 \ell_3} = \sum_{m_i} \(\begin{array}{ccc}\ell_1 & \ell_2 & \ell_3 \\ m_1 & m_2 & m_3  \end{array}\) \sqrt{\frac{v_{\ell_1}v_{\ell_2}v_{\ell_3}}{C^{TT}_{\ell_1} C^{TT}_{\ell_2} C^{TT}_{\ell_3}}}\bigg(a^{T}_{\ell_1 m_1} a^{T}_{\ell_2 m_2} a^{T}_{\ell_3 m_3} - \<a^{T}_{\ell_1 m_1} a^{T}_{\ell_2 m_2}\> a^{T}_{\ell_3 m_3} - \<a^{T}_{\ell_2 m_2} a^{T}_{\ell_3 m_3}\> a^{T}_{\ell_1 m_1} - \<a^{T}_{\ell_1 m_1} a^{T}_{\ell_3 m_3}\> a^{T}_{\ell_2 m_2}\bigg)\,,
\end{align}
then we can decompose this into the CMB basis as well. Thus we define
\begin{align}
\beta^{TTT}_{n} = \sum_p\gamma^{-1}_{np} \<Q_p,\, S^{TTT}_{D}\>_\ell\,.
\end{align}
The estimator then takes the simple form
\begin{align}
\curl{E} &= \frac{\sum_{mn} \alpha^{TTT}_m \gamma_{mn}\, \beta^{TTT}_n}{\sum_{mn} \alpha^{TTT}_m \gamma_{mn}\, \alpha^{TTT}_n}\,.
\end{align}
While this definition has nice symmetry it is in fact easier to work with $\hat{\beta}_n = \sum_p \gamma_{np} \beta_p = \<Q_p,\, S^{TTT}_{D}\>_\ell$ as this is simpler to calculate in practice.
If we transform to the orthonormal basis $R_n$, making the definition $\beta^R_n = \sum_p \lambda^{-1}_{pn} \beta_p  = \sum_p \lambda_{pn} \hat{\beta}_n$, then the estimator is
\begin{align}
\curl{E} &= \frac{\sum_n \alpha^{RTTT}_n \beta^{RTTT}_n}{\sum_n {\alpha^{RTTT}_n}^2}\,.
\end{align}
This has two very usefull properties. The first is that the normalisation is very easy to calculate and the second is that if the data contains a bispectrum described by mode vector $\alpha^R$ then $\<\beta^R\> = \alpha^R$.

Now we will define our primordial and late time basis to be symmetric combinations of separable functions which will allow us to simplify the calculations required to form the estimator.  We make the general definitions
\begin{align}
\nonumber \bar{Q}_n(\klist) &= \frac{1}{6} \( \bar{q}_i(k_1)\bar{q}_j(k_2)\bar{q}_k(k_3) + \bar{q}_j(k_1)\bar{q}_k(k_2)\bar{q}_i(k_3) + \bar{q}_k(k_1)\bar{q}_i(k_2)\bar{q}_j(k_3) + \bar{q}_k(k_1)\bar{q}_j(k_2)\bar{q}_i(k_3) + \bar{q}_j(k_1)\bar{q}_i(k_2)\bar{q}_k(k_3) \right. \\ &+ \left.  \bar{q}_i(k_1)\bar{q}_k(k_2)\bar{q}_j(k_3)\)\,,\\
Q^{TTT}_{n\,l_1 l_2 l_3} &= \frac{1}{6} \( q_{il_1}q_{jl_2}q_{kl_3} + q_{jl_1}q_{kl_2}q_{il_3} + q_{kl_1}q_{il_2}q_{jl_3} + q_{kl_1}q_{jl_2}q_{il_3} + q_{jl_1}q_{il_2}q_{kl_3} + q_{il_1}q_{kl_2}q_{jl_3}\)\,.
\end{align}
The $\bar{q}$ and $q$ can be any arbitrary functions.  Commonly in the modal method we work with either polynomials or trigonometric functions, as in the Planck analysis, or trigonometric functions of logarithms, which are currently in development for resonance type models. Clearly there are many other reasonable choices. Two obvious examples are: top hat functions, to localise the signal in $\ell$-space, which creates a Binned estimator; or the harmonic transform of wavelets, to localise the signal in real space, to create a wavelet estimator. While the localisation is clearly desirable these methods have their drawbacks. While the binned estimator is automatically orthogonal it requires many more modes to represent the same data (over 10000 compared to our ~500 for the scale invariant shapes). Although less modes are required for wavelets the individual $Q$ are highly correlated so $\gamma$ is very close to singular. Inversion requires techniques like Principle Component Analysis (PCA) to remove degenerate combinations and to stabilise the method.

The above expressions require a mapping between the mode number $n$ and the $ijk$ triple identifying the combination of the underlying $q$.  This is again a free choice. We frequently take a distance ordering where the modes are ordered first by increasing $i^2+j^2+k^2$ (with $i \le j \le k$) then for degenerate modes by increasing $k$ then $j$. This leads to mappings of the form
\begin{align}
\begin{array}{lclll}
n & \rightarrow & i & j & k\\
\hline
0 & \rightarrow & 0 & 0 & 0\\
1 & \rightarrow & 0 & 0 & 1\\
2 & \rightarrow & 0 & 1 & 1\\
3 & \rightarrow & 1 & 1 & 1\\
4 & \rightarrow & 0 & 0 & 2\\
5 & \rightarrow & 0 & 1 & 2\\
\cdots
\end{array}
\end{align}
Now we have defined our basis we can write explicit expressions for calculation of $\Gamma$, $\gamma$ and $\beta$ (the calculation of the initial $\bar{\alpha}$ and $\bar{\gamma}$ enjoy no simplification and must be done as full 3D calculations). For large numbers of basis functions the $\gamma$ and $\bar{\gamma}$ matrices can become numerically degenerate and so the basis becomes unstable.  We overcome this by considering a larger set of modes than required then iteratively removing the mode which leave a (unit normalised) $\gamma$ matrix with the greatest minimum-eigenvalue. This can be done until the minimum-eigenvalue of $\gamma$ is above some threshold, say $10^{-6}$, producing a stable basis. We define the projected version of our individual primordial basis functions
\begin{align}
\tilde{q}_{i\ell}(x) = \sqrt{\frac{v^2_\ell}{C^{TT}_{\ell}}} \int dk \bar{q}_i(k) \Delta^T_\ell(k) \, j_\ell(kx)\,,
\end{align}
which give $\widetilde{Q}$ in the simple form
\begin{align}
\widetilde{Q}^{TTT}_{n} &= \frac{1}{6} \int x^2dx \(\tilde{q}_{i\ell_1}(x) \tilde{q}_{j\ell_2}(x) \tilde{q}_{k\ell_3}(x) + 5\,\mbox{perms}\)\,.
\end{align}
Now we can write our orthogonalisation and projection matrices, $\gamma$ and $\Gamma$, in similarly simple form. To do this we must note that we can write
\begin{align}
h^2_{\ell_1 \ell_2 \ell_3} = \frac{(2\ell_1+1)(2\ell_2+1)(2\ell_3+1)}{8\pi} \int d\mu \,P_{\ell_1}(\mu)\, P_{\ell_2}(\mu)\, P_{\ell_3}(\mu)\,,
\end{align}
which can be calculated exactly with Gauss-Legendre integration. We also define the two $\ell$ sums
\begin{align}
P_{ii'}(\mu) &\equiv \sum_\ell \frac{(2\ell+1)}{v^2_\ell} q_{i\ell} q_{i'\ell} P_\ell(\mu)\,,\\
\widetilde{P}_{ii'}(x,\mu) &\equiv \sum_\ell \frac{(2\ell+1)}{v^2_\ell} \tilde{q}_{i\ell}(x) q_{i'\ell} P_\ell(\mu)\,.
\end{align}
Then $\gamma$ and $\Gamma$ then take the simple form
\begin{align}
\gamma_{mn} &= \frac{1}{48\pi} \int d\mu \(P_{ii'}(\mu)P_{jj'}(\mu)P_{kk'}(\mu) + 5\,\mbox{perms}\)\,,\\
\Gamma_{mn} &= \frac{1}{48\pi} {\gamma_{mn'}}^{-1} \int d\mu \int x^2 dx \(\widetilde{P}_{ii'}(x,\mu) \widetilde{P}_{jj'}(x,\mu) \widetilde{P}_{kk'}(x,\mu) + 5\,\mbox{perms}\)\,,
\end{align}
where the permutations are only over the primed indices. The $\hat{\beta}$ can be formed from the product of filtered maps
\begin{align}
\hat{\beta}_{n} &= \int d\Omega_n \(M_{i}(\hat{n}) M_{j}(\hat{n})M_{k}(\hat{n}) - 3 {\bold M}_{ij}(\hat{n}) M_{k}(\hat{n})\)\,,
\end{align}
where we have defined the filtered map $M$ and its covariance by
\begin{align}
M_{i}(\hat{n}) \equiv \sum_{\ell m} \frac{1}{\sqrt{v^2_{\ell}C^{TT}_{\ell}}} q_{i\ell} a_{\ell m} Y_{\ell m}(\hat{n})\,,\quad & {\bold M}_{ij}(\hat{n}) = \<M_{i}(\hat{n})M_{j}(\hat{n})\>\,.
\end{align}

We can see that the KSW estimator is a special case of the modal estimator. If we define our primordial basis to be $\bar{q}_i(k) = k^{i-1}$ where $i=(0,1,2,3)$ and use the three possible scale invariant combinations of them to form $\bar{Q}$ with the ordering
\begin{align}
\begin{array}{lclll}
n & \rightarrow & i & j & k\\
\hline
0 & \rightarrow & 1 & 1 & 1\\
1 & \rightarrow & 0 & 1 & 2\\
2 & \rightarrow & 0 & 0 & 3
\end{array}
\end{align}
Then by taking the $\widetilde{Q}$ to be our late time basis, $Q$, we obtain the KSW estimator. Here no orthogonalisation is required. We can read the $\alpha\, (= \bar{\alpha})$ coefficients directly from the theoretical templates so
\begin{align}
\alpha^{local} &= 6A^2\(0,0,1\)\,,\\
\alpha^{equil} &= 6A^2\(-2,6,-3\)\,,\\
\alpha^{ortho} &= 6A^2\(-8,18,-9\)\,.
\end{align}
and $A$ is the amplitude of the primordial power spectrum. To put this in the usual KSW notation
\begin{align}
\tilde{q}_{0\ell}(x) = \beta_\ell(x)\,, \quad \tilde{q}_{1\ell}(x) = \delta_\ell(x)\,, \quad \tilde{q}_{2\ell}(x) = \gamma_\ell(x)\,, \quad \tilde{q}_{3\ell}(x) = \alpha_\ell(x)
\end{align}
This approach has the benefit that it is exact (up to approximating the model by with a template) and so provides a good benchmark for comparing methods. However as we have used $\widetilde{Q}$ as our late time basis we now have an extra integral over $x$ in our calculation of $\hat{\beta}$. Hence this method is much slower which can be problematic when there are large numbers of maps to analyse. The method is obviously also limited to these three templates. It cannot tackle general models unless suitable separable approximations in terms of well behaved functions can be found.

\section{Extension to include polarisation}

We now consider the case where we wish to include the information from $E$-mode polarisation into the estimator. Now instead of just $TTT$ we have 8 CMB bispectra to consider and must calculate the contribution from all possible pairings of theory and data. We will take the estimator, \eqref{eq:estimator}, and rewrite it in matrix form so the full complexity is visible.  We define the theory and data vectors $B$ and $A$ and the inverse covariance matrix $C$ as follows

\begin{align}
B &= \[\begin{array}{c}
b^{TTT}_{\ell_1 \ell_2 \ell_3} \\
b^{TTE}_{\ell_1 \ell_2 \ell_3} \\
b^{TET}_{\ell_1 \ell_2 \ell_3} \\
b^{ETT}_{\ell_1 \ell_2 \ell_3} \\
b^{TEE}_{\ell_1 \ell_2 \ell_3} \\
b^{ETE}_{\ell_1 \ell_2 \ell_3} \\
b^{EET}_{\ell_1 \ell_2 \ell_3} \\
b^{EEE}_{\ell_1 \ell_2 \ell_3}
\end{array}\]
\,,\quad A = \[\begin{array}{c}
a_{\ell_1 m_1}^T a_{\ell_2 m_2}^T a_{\ell_3 m_3}^T - \<a^{T}_{\ell_1 m_1} a^{T}_{\ell_2 m_2}\> a^{T}_{\ell_3 m_3} - \<a^{T}_{\ell_2 m_2} a^{T}_{\ell_3 m_3}\> a^{T}_{\ell_1 m_1} - \<a^{T}_{\ell_1 m_1} a^{T}_{\ell_3 m_3}\> a^{T}_{\ell_2 m_2}\\
a_{\ell_1 m_1}^T a_{\ell_2 m_2}^T a_{\ell_3 m_3}^E - \<a^{T}_{\ell_1 m_1} a^{T}_{\ell_2 m_2}\> a^{E}_{\ell_3 m_3} - \<a^{T}_{\ell_2 m_2} a^{E}_{\ell_3 m_3}\> a^{T}_{\ell_1 m_1} - \<a^{T}_{\ell_1 m_1} a^{E}_{\ell_3 m_3}\> a^{T}_{\ell_2 m_2} \\
a_{\ell_1 m_1}^T a_{\ell_2 m_2}^E a_{\ell_3 m_3}^T - \<a^{T}_{\ell_1 m_1} a^{E}_{\ell_2 m_2}\> a^{T}_{\ell_3 m_3} - \<a^{E}_{\ell_2 m_2} a^{T}_{\ell_3 m_3}\> a^{T}_{\ell_1 m_1} - \<a^{T}_{\ell_1 m_1} a^{T}_{\ell_3 m_3}\> a^{E}_{\ell_2 m_2} \\
a_{\ell_1 m_1}^E a_{\ell_2 m_2}^T a_{\ell_3 m_3}^T - \<a^{E}_{\ell_1 m_1} a^{T}_{\ell_2 m_2}\> a^{T}_{\ell_3 m_3} - \<a^{T}_{\ell_2 m_2} a^{T}_{\ell_3 m_3}\> a^{E}_{\ell_1 m_1} - \<a^{E}_{\ell_1 m_1} a^{T}_{\ell_3 m_3}\> a^{T}_{\ell_2 m_2} \\
a_{\ell_1 m_1}^T a_{\ell_2 m_2}^E a_{\ell_3 m_3}^E - \<a^{T}_{\ell_1 m_1} a^{E}_{\ell_2 m_2}\> a^{E}_{\ell_3 m_3} - \<a^{E}_{\ell_2 m_2} a^{E}_{\ell_3 m_3}\> a^{T}_{\ell_1 m_1} - \<a^{T}_{\ell_1 m_1} a^{E}_{\ell_3 m_3}\> a^{E}_{\ell_2 m_2} \\
a_{\ell_1 m_1}^E a_{\ell_2 m_2}^T a_{\ell_3 m_3}^E - \<a^{E}_{\ell_1 m_1} a^{T}_{\ell_2 m_2}\> a^{E}_{\ell_3 m_3} - \<a^{T}_{\ell_2 m_2} a^{E}_{\ell_3 m_3}\> a^{E}_{\ell_1 m_1} - \<a^{E}_{\ell_1 m_1} a^{E}_{\ell_3 m_3}\> a^{T}_{\ell_2 m_2} \\
a_{\ell_1 m_1}^E a_{\ell_2 m_2}^E a_{\ell_3 m_3}^T - \<a^{E}_{\ell_1 m_1} a^{E}_{\ell_2 m_2}\> a^{T}_{\ell_3 m_3} - \<a^{E}_{\ell_2 m_2} a^{T}_{\ell_3 m_3}\> a^{E}_{\ell_1 m_1} - \<a^{E}_{\ell_1 m_1} a^{T}_{\ell_3 m_3}\> a^{E}_{\ell_2 m_2} \\
a_{\ell_1 m_1}^E a_{\ell_2 m_2}^E a_{\ell_3 m_3}^E - \<a^{E}_{\ell_1 m_1} a^{E}_{\ell_2 m_2}\> a^{E}_{\ell_3 m_3} - \<a^{E}_{\ell_2 m_2} a^{E}_{\ell_3 m_3}\> a^{E}_{\ell_1 m_1} - \<a^{E}_{\ell_1 m_1} a^{E}_{\ell_3 m_3}\> a^{E}_{\ell_2 m_2}
\end{array}\]\,,\\
C &= \[\begin{array}{cccccccc}
T_{\ell_1} T_{\ell_2} T_{\ell_3} & T_{\ell_1} T_{\ell_2} M_{\ell_3} & T_{\ell_1} M_{\ell_2} T_{\ell_3} & M_{\ell_1} T_{\ell_2} T_{\ell_3} & T_{\ell_1} M_{\ell_2} M_{\ell_3} & M_{\ell_1} T_{\ell_2} M_{\ell_3} & M_{\ell_1} M_{\ell_2} T_{\ell_3} & M_{\ell_1} M_{\ell_2} M_{\ell_3} \\
T_{\ell_1} T_{\ell_2} M_{\ell_3} & T_{\ell_1} T_{\ell_2} E_{\ell_3} & M_{\ell_1} M_{\ell_2} T_{\ell_3} & M_{\ell_1} T_{\ell_2} M_{\ell_3} & T_{\ell_1} M_{\ell_2} E_{\ell_3} & M_{\ell_1} T_{\ell_2} E_{\ell_3} & M_{\ell_1} M_{\ell_2} M_{\ell_3} & M_{\ell_1} M_{\ell_2} E_{\ell_3} \\
T_{\ell_1} M_{\ell_2} T_{\ell_3} & T_{\ell_1} M_{\ell_2} M_{\ell_3} & T_{\ell_1} E_{\ell_2} T_{\ell_3} & M_{\ell_1} M_{\ell_2} T_{\ell_3} & T_{\ell_1} E_{\ell_2} M_{\ell_3} & M_{\ell_1} M_{\ell_2} M_{\ell_3} & M_{\ell_1} E_{\ell_2} T_{\ell_3} & M_{\ell_1} E_{\ell_2} M_{\ell_3} \\
M_{\ell_1} T_{\ell_2} T_{\ell_3} & M_{\ell_1} T_{\ell_2} M_{\ell_3} & T_{\ell_1} M_{\ell_2} M_{\ell_3} & E_{\ell_1} T_{\ell_2} T_{\ell_3} & M_{\ell_1} M_{\ell_2} M_{\ell_3} & E_{\ell_1} T_{\ell_2} M_{\ell_3} & E_{\ell_1} M_{\ell_2} T_{\ell_3} & E_{\ell_1} M_{\ell_2} M_{\ell_3} \\
T_{\ell_1} M_{\ell_2} M_{\ell_3} & T_{\ell_1} M_{\ell_2} E_{\ell_3} & T_{\ell_1} E_{\ell_2} M_{\ell_3} & M_{\ell_1} M_{\ell_2} M_{\ell_3} & T_{\ell_1} E_{\ell_2} E_{\ell_3} & M_{\ell_1} M_{\ell_2} E_{\ell_3} & M_{\ell_1} E_{\ell_2} M_{\ell_3} & M_{\ell_1} E_{\ell_2} E_{\ell_3} \\
M_{\ell_1} T_{\ell_2} M_{\ell_3} & M_{\ell_1} T_{\ell_2} E_{\ell_3} & M_{\ell_1} M_{\ell_2} M_{\ell_3} & E_{\ell_1} T_{\ell_2} M_{\ell_3} & M_{\ell_1} M_{\ell_2} E_{\ell_3} & E_{\ell_1} T_{\ell_2} E_{\ell_3} & E_{\ell_1} M_{\ell_2} M_{\ell_3} & E_{\ell_1} M_{\ell_2} E_{\ell_3} \\
M_{\ell_1} M_{\ell_2} T_{\ell_3} & M_{\ell_1} M_{\ell_2} M_{\ell_3} & M_{\ell_1} E_{\ell_2} T_{\ell_3} & E_{\ell_1} M_{\ell_2} T_{\ell_3} & M_{\ell_1} E_{\ell_2} M_{\ell_3} & E_{\ell_1} M_{\ell_2} M_{\ell_3} & E_{\ell_1} E_{\ell_2} T_{\ell_3} & E_{\ell_1} E_{\ell_2} M_{\ell_3} \\
M_{\ell_1} M_{\ell_2} M_{\ell_3} & M_{\ell_1} M_{\ell_2} E_{\ell_3} & M_{\ell_1} E_{\ell_2} M_{\ell_3} & E_{\ell_1} M_{\ell_2} M_{\ell_3} & M_{\ell_1} E_{\ell_2} E_{\ell_3} & E_{\ell_1} M_{\ell_2} E_{\ell_3} & E_{\ell_1} E_{\ell_2} M_{\ell_3} & E_{\ell_1} E_{\ell_2} E_{\ell_3}
\end{array}\]\,, \label{eq:corrmat}
\end{align}
where we have make the definitions
\begin{align}
T_{\ell} = (C^{-1})^{TT}_{\ell} \px \frac{C^{EE}_{\ell}}{C^{TT}_{\ell} C^{EE}_{\ell} - {C^{TE}_{\ell}}^2}\,,\quad
E_{\ell} = (C^{-1})^{EE}_{\ell} \px \frac{C^{TT}_{\ell}}{C^{TT}_{\ell} C^{EE}_{\ell} - {C^{TE}_{\ell}}^2}\,,\quad
M_{\ell} = (C^{-1})^{TE}_{\ell} \px \frac{-C^{TE}_{\ell}}{C^{TT}_{\ell} C^{EE}_{\ell} - {C^{TE}_{\ell}}^2}\,.
\end{align}
The estimator then takes the form
\begin{align}\label{fullTEest}
\curl{E} = \frac{\sum_{\ell_i m_i} \curl{G}^{\ell_1 \ell_2 \ell_3}_{m_1 m_2 m_3} B^T C A}{\sum_{l_i} h^2_{\ell_1 \ell_2 \ell_3} B^T C B}\,.
\end{align}
This is much more complicated than the temperature only case, requiring the calculation of 13 unique terms for both the estimator and its normalisation. It is also unclear what we should define as the shape function as all eight bispectra are mixed. It is possible to apply the modal method directly to \eqref{fullTEest}, and this will be the focus of a companion paper \cite{LSFinprep}, but here we will describe a novel solution.

The complexity in this calculation is entirely down to the coupling between $T$ and $E$ as all off diagonal terms contain a mixing part $M_{\ell}$. We pause to note that this is the simplest possible case.  If we wished to consider the trispectrum with both $T$ and $E$ there would be 16 coupled trispectra producing 22 unique terms. The obvious solution is to find a way to decouple the data sets and instead to work with an orthogonalised set of $a_{lm}$.  This is achieved by taking the covariance matrix, \eqref{eq:covariance}, calculating its inverse then performing a Cholesky decomposition, $C^{-1} = L L^T$, (which as mentioned previously is equivalent to performing a modified Gramm-Schmidt orthogonalisation). The decomposition yields
\begin{align}\label{eq:cholesky}
L = \(\begin{array}{cc}
\frac{1}{\sqrt{C^{TT}_\ell}} & 0 \\
\frac{-C^{TE}_\ell}{\sqrt{C^{TT}_\ell}\sqrt{C^{TT}_\ell C^{EE}_\ell - {C^{TE}_\ell}^2}} & \frac{C^{TT}_\ell}{\sqrt{C^{TT}_\ell}\sqrt{C^{TT}_\ell C^{EE}_\ell - {C^{TE}_\ell}^2}}
\end{array}\)\,.
\end{align}
So if we define new $a_{\ell m}$ such that
\begin{align}
\[\begin{array}{c} \hat{a}^T_{\ell m} \\ \hat{a}^E_{\ell m}\end{array}\] =
\[\begin{array}{c}
\frac{a^T_{\ell m}}{\sqrt{C^{TT}_\ell}} \\
\frac{C^{TT}_\ell a^E_{\ell m} - C^{TE}_\ell a^T_{\ell m}}{\sqrt{C^{TT}_\ell}\sqrt{C^{TT}_\ell C^{EE}_\ell - {C^{TE}_\ell}^2}}
\end{array}\]\,,
\end{align}
then these new $\hat{a}_{\ell m}$ are orthonormal and so there are no cross terms to consider. We note here that this is equivalent to orthonormalising the bispectra by performing a Cholesky decomposition on the correlation matrix $C$ given by \eqref{eq:corrmat}. The order in which the bispectra are orthonormalised are
\begin{align}\label{eq:biorder}
TTT \rightarrow TTE \rightarrow TEE \rightarrow EEE\,.
\end{align}
Armed with this we can generalise the expression for the CMB shape function, \eqref{eq:CMBshape}, to include polarisation
\begin{align}
S^{X_1X_2X_3}_{\ell_1 \ell_2 \ell_3} \equiv \frac{v_{\ell_1}v_{\ell_2}v_{\ell_3}}{h_{\ell_1 \ell_2 \ell_3}} \sum_{m_i} \(\begin{array}{ccc}\ell_1 & \ell_2 & \ell_3 \\ m_1 & m_2 & m_3  \end{array}\) \<\hat{a}^{X_1}_{\ell_1 m_1}\hat{a}^{X_2}_{\ell_2 m_2}\hat{a}^{X_3}_{\ell_3 m_3}\>\,,
\end{align}
which reduces to the previous expression for $S^{TTT}$.  Now we need to define multiple late time basis taking into account the symmetries of the bispectra being decomposed as
\begin{align}
Q^{XXX}_{n\,\ell_1 \ell_2 \ell_3} &= \frac{1}{6} \( q^{X}_{i\ell_1}q^{X}_{j\ell_2}q^{X}_{k\ell_3} + q^{X}_{j\ell_1}q^{X}_{k\ell_2}q^{X}_{i\ell_3} + q^{X}_{k\ell_1}q^{X}_{i\ell_2}q^{X}_{j\ell_3} + q^{X}_{k\ell_1}q^{X}_{j\ell_2}q^{X}_{i\ell_3} + q^{X}_{j\ell_1}q^{X}_{i\ell_2}q^{X}_{k\ell_3} + q^{X}_{i\ell_1}q^{X}_{k\ell_2}q^{X}_{j\ell_3}\)\,,\\
Q^{XXY}_{n\,\ell_1 \ell_2 \ell_3} &= \frac{1}{2} \( q^{X}_{i\ell_1}q^{X}_{j\ell_2}q^{Y}_{k\ell_3} + q^{X}_{j\ell_1}q^{X}_{i\ell_2}q^{Y}_{k\ell_3}\)\,,\\
Q^{XYY}_{n\,\ell_1 \ell_2 \ell_3} &= \frac{1}{2} \( q^{X}_{i\ell_1}q^{Y}_{j\ell_2}q^{Y}_{k\ell_3} + q^{X}_{i\ell_1}q^{Y}_{k\ell_2}q^{Y}_{j\ell_3}\)\,,
\end{align}
where we have allowed the possibility of using different basis functions, $q^X_i(\ell)$, in $\ell$-space for temperature and E- and B-mode polarisation. We must also define new mappings between the mode number $n$ and the $ijk$ triple reflecting the differing symmetries of the $Q$.  We again take a distance ordering where the modes are ordered first by increasing $i^2+j^2+k^2$ (with $i \le j \le k$ for $Q^{XXX}$; $i<j$ for $Q^{XXY}$; and $j<k$ for $Q^{XYY}$) and for degenerate modes by increasing $k$ then $j$. This leads to mappings of the form
\begin{align}
\begin{array}{c|c|c}
Q_n^{XXX} & Q_n^{XXY} & Q_n^{XYY}\\
\hline
\begin{array}{lclll}
n & \rightarrow & i & j & k\\
\hline
0 & \rightarrow & 0 & 0 & 0\\
1 & \rightarrow & 0 & 0 & 1\\
2 & \rightarrow & 0 & 1 & 1\\
3 & \rightarrow & 1 & 1 & 1\\
4 & \rightarrow & 0 & 0 & 2\\
5 & \rightarrow & 0 & 1 & 2\\
\cdots
\end{array}
&
\begin{array}{lclll}
n & \rightarrow & i & j & k\\
\hline
0 & \rightarrow & 0 & 0 & 0\\
1 & \rightarrow & 0 & 1 & 0\\
2 & \rightarrow & 0 & 0 & 1\\
3 & \rightarrow & 1 & 1 & 0\\
4 & \rightarrow & 0 & 1 & 1\\
5 & \rightarrow & 1 & 1 & 1\\
\cdots
\end{array}
&
\begin{array}{lclll}
n & \rightarrow & i & j & k\\
\hline
0 & \rightarrow & 0 & 0 & 0\\
1 & \rightarrow & 0 & 0 & 1\\
2 & \rightarrow & 1 & 0 & 0\\
3 & \rightarrow & 0 & 1 & 1\\
4 & \rightarrow & 1 & 0 & 1\\
5 & \rightarrow & 1 & 1 & 1\\
\cdots
\end{array}
\end{array}\,.
\end{align}

Note that where basis functions lack full symmetry we require more modes to obtain the same resolution.  This is because for modes with one axis of symmetry, of the form $Q^{XXY}$ and $Q^{XYY}$, you only need to fit half of the full domain; and for modes with full symmetry, with the form $Q^{XXX}$, only require a sixth. Thus we may need to consider a larger number of basis functions to compensate for the loss of resolution when cross correlating data sets.

We now define the orthonormalised transfer functions
\begin{align}
\hat{\Delta}^T_\ell(k) &\equiv \frac{\Delta^T_\ell(k)}{\sqrt{C^{TT}_\ell}}\,,\\
\hat{\Delta}^E_\ell(k) &\equiv \frac{C^{TT}_\ell\Delta^E_\ell(k) - C^{TE}_\ell\Delta^T_\ell(k)}{\sqrt{C^{TT}_\ell}\sqrt{C^{TT}_\ell C^{EE}_\ell - {C^{TE}_\ell}^2}}\,,
\end{align}
which allows us to now extend the projected primordial basis $\widetilde{Q}_n$, \eqref{eq:projbasis}, to include polarisation. We want the previous relation between primordial and late time shape functions to continue to hold, namely that
\begin{align}
\bar{S}(\klist) = \sum_n \bar{\alpha}_n\,\bar{Q}_n(\klist) \rightarrow
S_{l_1 l_2 l_3} = \sum_n \bar{\alpha}_n \,\widetilde{Q}_{n\,l_1 l_2 l_3}\,.
\end{align}
The $\widetilde{Q}$ which satisfy this have the same simple form we had previously, allowing for the new symmetries,
\begin{align}
\widetilde{Q}^{XXX}_{mn} &= \frac{1}{6} \int x^2dx \(\tilde{q}^{X}_{i\ell_1}(x) \tilde{q}^{X}_{j\ell_2}(x) \tilde{q}^{X}_{k\ell_3}(x) + 5\,\mbox{perms}\)\,,\\
\widetilde{Q}^{XXY}_{mn} &= \frac{1}{2} \int x^2dx \(\tilde{q}^{X}_{i\ell_1}(x) \tilde{q}^{X}_{j\ell_2}(x) \tilde{q}^{Y}_{k\ell_3}(x) + \tilde{q}^{X}_{j\ell_1}(x) \tilde{q}^{X}_{i\ell_2}(x) \tilde{q}^{Y}_{k\ell_3}(x)\)\,,\\
\widetilde{Q}^{XYY}_{mn} &= \frac{1}{2} \int x^2dx \(\tilde{q}^{X}_{i\ell_1}(x) \tilde{q}^{Y}_{j\ell_2}(x) \tilde{q}^{Y}_{k\ell_3}(x) + \tilde{q}^{X}_{i\ell_1}(x) \tilde{q}^{Y}_{k\ell_2}(x) \tilde{q}^{Y}_{j\ell_3}(x)\)\,.
\end{align}
Here we have made the obvious extension to the definition of $\tilde{q}$ by replacing the transfer function with its orthogonalised version
\begin{align}\label{eq:qtildenew}
\tilde{q}^{X}_{i\ell}(x) = v_\ell \int dk \bar{q}_i(k) \hat{\Delta}^X_\ell(k) \, j_\ell(kx)\,.
\end{align}
The remaining equations for the orthogonalisation matrix $\gamma$ and projection matrix $\Gamma$ are now unchanged
\begin{align}
\gamma^{XYZ}_{mn} &= \frac{1}{48\pi} \int d\mu \(P^X_{ii'}(\mu)P^Y_{jj'}(\mu)P^Z_{kk'}(\mu) + 5\,\mbox{perms}\)\,,\\
\nonumber \Gamma^{XYZ}_{mn} &= \frac{1}{48\pi} {\gamma^{XYZ}_{mn'}}^{-1} \int d\mu \int x^2 dx \(\widetilde{P}^X_{ii'}(x,\mu)\widetilde{P}^Y_{jj'}(x,\mu)\widetilde{P}^Z_{kk'}(x,\mu) + 5\,\mbox{perms}\)\,.
\end{align}
This allows us to calculate all sets of $\alpha^{XYZ}$ from the single $\bar{\alpha}$ needed for the estimator. The data coefficients $\hat{\beta}^{XYZ}$ simplify in a similar way so that
\begin{align}
\hat{\beta}^{XYZ}_{n} &= \int d\Omega_n M^X_{i}(\hat{n}) M^Y_{j}(\hat{n})M^Z_{k}(\hat{n}) - {\bold M}^{XY}_{ij}(\hat{n}) M^Z_{k}(\hat{n}) - {\bold M}^{XZ}_{ik}(\hat{n}) M^Y_{j}(\hat{n}) - {\bold M}^{YZ}_{jk}(\hat{n}) M^Z_{i}(\hat{n})\,,
\end{align}
where we have defined the new filtered maps replacing the multipoles with there orthogonalised counterparts
\begin{align}\label{eq:mapsnew}
M^X_{i}(\hat{n}) \equiv \sum_{lm} \frac{1}{v_{\ell}} q^X_{i \ell} \hat{a}^X_{\ell m} Y_{\ell m}(\hat{n})\quad & {\bold M}^{XY}_{ij}(\hat{n}) = \<M^X_{i}(\hat{n})M^Y_{j}(\hat{n})\>\,.
\end{align}
Finally the estimator becomes
\begin{align}
\curl{E} &= \frac{\sum_{X_i}\sum_{n} \alpha^{RX_1X_2X_3}_n \beta^{RX_1X_2X_3}_n}{\sum_{X_i}\sum_n {\alpha^{RX_1X_2X_3}_n}^2} = \frac{\sum_{n} \alpha^{RTTT}_n \beta^{RTTT}_n + 3\sum_{n} \alpha^{RTTE}_n \beta^{RTTE}_n + 3\sum_{n} \alpha^{RTEE}_n \beta^{RTEE}_n + \sum_{n} \alpha^{REEE}_n \beta^{REEE}_n}{\sum_n {\alpha^{RTTT}_n}^2 + 3\sum_n {\alpha^{RTTE}_n}^2 + 3\sum_n {\alpha^{RTEE}_n}^2 + \sum_n {\alpha^{REEE}_n}^2}\,,
\end{align}
where $X_i$ runs over all combinations of $T$ and $E$. As noted earlier the orthonormalisation of the $a_{\ell m}$ is equivalent to the orthonormalisation of the weighted bispectra themselves. Hence the sum over $X_i$ can also be terminated at any point provided all higher terms in the orthogonalisation, as defined by \eqref{eq:biorder}, are included. This format allows us to trivially switch between the $T$-only estimator, the $T+E$ estimator and any allowed T plus partial polarisation estimators.

All of the above has been calculated in the diagonal covariance approximation which has been shown to be near-optimal for Planck temperature data when in-painting is applied.  The reader may be concerned that this assumption may not hold now we have extended the analysis to include polarisation.  We note two easy ways for the method to be adapted if this approximation breaks down.  The first is that if the inverse covariance weighted maps can be calculated these can be used in this framework by making the substitution of $a^X_{\ell m} \rightarrow C^X_\ell C^{-1}(a^X_{\ell m})$. The second is that the $\beta$ covariance matrices can be calculated from realistic simulations and the estimator then becomes
\begin{align}
\curl{E} &= \frac{\sum_{X^{\phantom{'}}_i,\,X'_i}\sum_{n} \alpha^{RX^{\phantom{'}}_1X^{\phantom{'}}_2X^{\phantom{'}}_3}_n \(C^{X^{\phantom{'}}_1X^{\phantom{'}}_2X^{\phantom{'}}_3,\,X'_1X'_2X'_3}_\beta\)^{-1} \beta^{RX'_1X'_2X'_3}_n}{\sum_{X^{\phantom{'}}_i,\,X'_i}\sum_n \alpha^{RX^{\phantom{'}}_1X^{\phantom{'}}_2X^{\phantom{'}}_3}_n \(C^{X^{\phantom{'}}_1X^{\phantom{'}}_2X^{\phantom{'}}_3,\,X'_1X'_2X'_3}_\beta\)^{-1}\alpha^{X'_1X'_2X'_3}_n}\,,
\end{align}
where $(C^{X^{\phantom{'}}_1X^{\phantom{'}}_2X^{\phantom{'}}_3,\,X'_1X'_2X'_3}_\beta)^{-1} $ is the inverse of the beta covariance matrix if the matrix is square, or Moore-Penrose pseudo inverse if rectangular (allowing for different numbers of modes for each bispectrum expanded). This is then the projected inverse covariance weighted estimator, as was shown in \cite{1105.2791}, which for a well chosen basis should be close to optimal.  Both approaches have their strengths. The first is optimal up to approximations used in calculating the Wiener filter for which the mask is modeled accurately, noise modeling is significantly simplified and no other effects are included. The second allows for including any realistic effect that can be simulated but will miss correlations which cannot be projected into the basis.  The extensions to the trispectrum are obvious and included as an appendix.  We note once again that while we have demonstraited the method with the modal formalism it applies equally to the KSW, binned and wavelet approaches.

\section{Simulations}
This method also allows the creation of simulations containing arbritrary small non-Gaussianity. Ordinarily this is complicated by the issue of modifying both temperature and polarisation so all cross correlations produce the correct result which is difficult.  As our rotated $a_{lm}$ are uncorrelated this is not an issue here and we can trivially extend the standard temperature-only method for separable shapes, first presented in \cite{0612571} then applied to the modal approach in \cite{0912.5516}. First we note that,
\begin{align}
\<\hat{a}^X_{\ell_1m_1}\hat{a}^Y_{\ell_2m_2}\> &= \delta_{\ell_1 \ell_2} \delta_{m_1 m_2} \delta_{XY}\,,
\end{align}
so $\hat{a}^X_{\ell m}$ are uncorrelated random variables with unit variance. This makes the creation of a Gaussian realisation, $(\hat{g}^T_{\ell m},\,\hat{g}^E_{\ell m})$, trivial. Then to introduce an small level of non-Gaussianity through a quadratic addition given by,
\begin{align}
\hat{a}^T_{\ell m} &= \hat{g}^T_{\ell m} + \frac{1}{6}\sum_{\ell_1 m_1 \ell_2 m_2} \frac{\int Y_{\ell_1 m_1} Y_{\ell_2 m_2} Y_{\ell m}}{v_{\ell_1} v_{\ell_2} v_\ell}\,f_{NL} \(S^{TTT}_{\ell_1 \ell_2 \ell}\,  \hat{g}^T_{\ell_1 m_1} \hat{g}^T_{\ell_2 m_2}\)\\
\hat{a}^E_{\ell m} &= \hat{g}^E_{\ell m} + \frac{1}{6}\sum_{\ell_1 m_1 \ell_2 m_2} \frac{\int Y_{\ell_1 m_1} Y_{\ell_2 m_2} Y_{\ell m}}{v_{\ell_1} v_{\ell_2} v_\ell}\, f_{NL}\(3 S^{TTE}_{\ell_1 \ell_2 \ell}\,  \hat{g}^T_{\ell_1 m_1} \hat{g}^T_{\ell_2 m_2} + 3 S^{TEE}_{\ell_1 \ell_2 \ell}\,  \hat{g}^T_{\ell_1 m_1} \hat{g}^E_{\ell_2 m_2} + S^{EEE}_{\ell_1 \ell_2 \ell}\,  \hat{g}^E_{\ell_1 m_1} \hat{g}^E_{\ell_2 m_2} \)\,.
\end{align}
Now we use $S = \sum \a_n Q_n$ to replace $S$ with its decomposition.  If we make the definitions,
\begin{align}
G^X_{i}(\hat{n}) \equiv \sum_{lm} \frac{1}{v_{\ell}} q^X_{i\ell} \hat{g}^X_{\ell m} Y_{\ell m}(\hat{n})\,, \quad {\bold G}^{X_1X_2}_{\ell m\,|\, ij} = \frac{1}{v_{\ell}}\int Y_{\ell m}(\un) G^{X_1}_{i}(\un) G^{X_2}_{j}(\un)\, d^2\un\,.
\end{align}
Then we have,
\begin{align}
\hat{a}^T_{\ell m} &= \hat{g}^T_{\ell m} + f_{NL} \sum_{n \rightarrow ijk} \frac{1}{18} \a^{TTT}_n\({\bold G}^{TT}_{ij} q^T_{k} + {\bold G}^{TT}_{jk} q^T_{i} + {\bold G}^{TT}_{ki} q^T_{j} \)_{\ell m} \\
\hat{a}^E_{\ell m} &= \hat{g}^E_{\ell m} + f_{NL} \sum_{n \rightarrow ijk} \(\frac{1}{2}\a^{TTE}_n {\bold G}^{TT}_{ij} q^E_{k} + \frac{1}{4}\a^{TEE}_n \({\bold G}^{TE}_{ij} q^E_{k} + {\bold G}^{TE}_{ik} q^E_{j}\) + \frac{1}{18}\a^{EEE}_n \({\bold G}^{EE}_{ij} q^E_{k} + {\bold G}^{EE}_{jk} q^E_{i} + {\bold G}^{EE}_{ki} q^E_{j}\) \)_{\ell m}\,.
\end{align}
Finally we can rotate back the the usual correlated multipoles,
\begin{align}
a^T_{\ell m} &= \sqrt{C^{TT}_\ell} \hat{a}^T_{\ell m}\\
a^E_{\ell m} &= \sqrt{C^{EE}_\ell - \frac{{C^{TE}_\ell}^2}{C^{TT}_\ell}}\, \hat{a}^E_{\ell m} - \sqrt{\frac{{C^{TE}_\ell}^2}{C^{TT}_\ell}}\, \hat{a}^T_{\ell m}\,.
\end{align}
These ideal maps can then undergo processing to include both real and experimental effects like lensing and noise. This presents a simple efficient method for creating temperature and polarisation simulations with small non-Gaussianity of arbitrary form.

\section{Conclusions}

Here we have presented a simple way to rewrite the optimal estimator for the bispectrum with temperature and polarisation CMB data. The new form significantly reduces the number of calculations required. In the simplest case, where we consider the bispectrum with just $T$ and $E$, the number of terms in the estimator is reduces by over a factor of three or by over four for the trispectrum.  The method was illustrated using the modal method but is applicable to any optimal approach including KSW, binned and wavelet methods. In addition to the great simplification of the equations this method also allows us calculate the correct signal-to-noise weight for each term.  This ensures that the convergence of the decompositions required for the modal, binned and wavelet methods is optimised.  The method has a straight-forward extension to higher order correlators like the trispectrum. It would also apply more generally to any situation where we need to consider multiple data sets with known correlations. One possible example would be calculating the galaxy bispectrum with multiple redshift bins.  It also presents a simple method for producing non-Gaussian simulations for any given model. The method is currently being implemented for the modal method and will be applied to the Planck  data with results appearing at the end of the year.

\section{Acknowledgements}

I am very grateful for many informative discussions with Michele Liguori and Paul Shellard. All numerical work was performed on the COSMOS supercomputer (an Altix 4700) part of the DIRAC HPC facility which is funded by STFC, HEFCE and SGI.  JRF was supported by STFC grant ST/F002998/1 and the Centre for Theoretical Cosmology.

\appendix
\section{Extension to Trispectrum}
The trispectrum in general depends on four vectors. Momentum conservation ensures the four vectors close and isotropy means that there is no dependence on the orientation of the resulting object. Thus the trispectrum only depends on the shape of a tetrahedron defined by the lenght of its six sides, the four vectors and two diagonals.  As the CMB is a two dimensional projection of the thee dimensional space the CMB trispectum is insensitive to one of the diagonals. The method applies to the full five dimensional case but here for simplicity we will work in the diagonal free approximation where the trispectrum only depends on the four lengths, $k_1,\,k_2,\,k_3,\,k_4$. The full five dimensional case can be easily deduced from the temperature only case described in \cite{1004.2915} and the discusion that follows.

The primordial and CMB trispectrum are defined by
\begin{align}
\< \Phi(\bk_1) \Phi(\bk_2) \Phi(\bk_3) \Phi(\bk_4)\> &= \(2\pi\)^3\delta(\bk_1+\bk_2+\bk_3+\bk_4) T(k_1,k_2,k_3,k_4)\,,\\
\< a^{X_1}_{\ell_1 m_1} a^{X_2}_{\ell_2 m_2} a^{X_3}_{\ell_3 m_3} a^{X_4}_{\ell_4 m_4} \> &= \curl{G}^{\ell_1 \ell_2 \ell_3 \ell_4}_{m_1 m_2 m_3 m_4} t^{X_1X_2X_3X_4}_{\ell_1 \ell_2 \ell_3 \ell_4}\,,
\end{align}
 and $\curl{G}$ is the 4D equivalent of the Gaunt integral defined as follows
\begin{align}
\curl{G}^{\ell_1 \ell_2 \ell_3 \ell_4}_{m_1 m_2 m_3 m_4} &= \int d\Omega_{\un} Y_{\ell_1 m_1 }(\un) Y_{\ell_2 m_2 }(\un) Y_{\ell_3 m_3 }(\un) Y_{\ell_4 m_4 }(\un) = \sum_{LM}(-1)^M\(\begin{array}{ccc}\ell_1 & \ell_2 & L \\ m_1 & m_2 & M  \end{array}\)\(\begin{array}{ccc}\ell_3 & \ell_4 & L \\ m_3 & m_4 & -M  \end{array}\) h_{\ell_1 \ell_2 L} h_{\ell_3 \ell_4 L}\,.
\end{align}

The optimal estimator for the trispectrum is \cite{1004.2915}
\begin{align}\label{eq:triestimator}
\curl{E} &= \frac{1}{N}\sum_{X_i X'_i}\sum_{\ell m}  \curl{G}^{\ell_1 \ell_2 \ell_3 \ell_4}_{m_1 m_2 m_3 m_4} t^{X_1X_2X_3X_4}_{\ell_1 \ell_2 \ell_3 \ell_4} (C^{-1})^{X_1X'_1}_{\ell_1} (C^{-1})^{X_2X'_2}_{\ell_2} (C^{-1})^{X_3X'_3}_{\ell_3} (C^{-1})^{X_4X'_4}_{\ell_4}\\
\nonumber & \[a^{X'_1}_{\ell_1 m_1} a^{X'_2}_{\ell_2 m_2} a^{X'_3}_{\ell_3 m_3} a^{X'_4}_{\ell_4 m_4} - \(\<a^{X'_1}_{\ell_1 m_1} a^{X'_2}_{\ell_2 m_2}\> a^{X'_3}_{\ell_3 m_3} a^{X'_4}_{\ell_4 m_4} + 5\mbox{perms}\) + \(\<a^{X'_1}_{\ell_1 m_1} a^{X'_2}_{\ell_2 m_2}\> \<a^{X'_3}_{\ell_3 m_3} a^{X'_4}_{\ell_4 m_4}\> + 2\mbox{perms}\) \]\,,
\end{align}
where $N$ is the normalisation
\begin{align}
N \equiv \sum_{X_i X'_i}\sum_{\ell_i}  \(\sum_L \frac{h^2_{\ell_1 \ell_2 L}h^2_{\ell_3 \ell_4 L}}{2L+1}\) t^{X_1 X_2 X_3 X_4}_{\ell_1 \ell_2 \ell_3 \ell_4} (C^{-1})^{X_1X'_1}_{\ell_1} (C^{-1})^{X_2X'_2}_{\ell_2} (C^{-1})^{X_3X'_3}_{\ell_3} (C^{-1})^{X_4X'_4}_{\ell_4} t^{X'_1 X'_2 X'_3 X'_4}_{\ell_1 \ell_2 \ell_3 \ell_4}\,.
\end{align}
The normalisation is related to the Fisher matrix by $N = 24F$. We can relate the primordial and CMB trispectra by
\begin{align}\label{eq:triprojection}
t^{X_1X_2X_3X_4}_{\ell_1 \ell_2 \ell_3 \ell_4} = \(\frac{2}{\pi}\)^4\int_{\curl{V}^T_k} \(k_1 k_2 k_3 k_4\)^2 T(k_1,k_2,k_3,k_4) \D^{X_1 X_2 X_3 X_4}_{\ell_1 \ell_2 \ell_3 \ell_4}(k_1,k_2,k_3,k_4)  d\curl{V}^T_k\,,
\end{align}
where we have defined
\begin{align}
\D^{X_1 X_2 X_3 X_4}_{\ell_1 \ell_2 \ell_3 \ell_4}(k_1,k_2,k_3,k_4) \equiv \D^{X_1}_{\ell_1}(k_1) \D^{X_2}_{\ell_2}(k_2) \D^{X_3}_{\ell_3}(k_3) \D^{X_4}_{\ell_4}(k_4) \int x^2 dx j_{\ell_1}(xk_1) j_{\ell_2}(xk_2) j_{\ell_3}(xk_3) j_{\ell_3}(xk_4)\,,
\end{align}
as the trispectrum transfer function. Using the above we can define the primordial shape function
\begin{align}
S(k_1,k_2,k_3,k_4) \equiv \(k_1 k_2 k_3 k_4\)^2 T(k_1,k_2,k_3,k_4)\,,
\end{align}
and the CMB shape function
\begin{align}
{S}^{X_1X_2X_3X_4}_{\ell_1 \ell_2 \ell_3 \ell_4} \equiv v_{\ell_1}v_{\ell_2}v_{\ell_3}v_{\ell_4} \sum_{m_i,M} (-1)^M\(\begin{array}{ccc}\ell_1 & \ell_2 & L \\ m_1 & m_2 & M  \end{array}\)\(\begin{array}{ccc}\ell_3 & \ell_4 & L \\ m_3 & m_4 & -M  \end{array}\) \<\hat{a}^{X_1}_{\ell_1 m_1}\hat{a}^{X_2}_{\ell_2 m_2}\hat{a}^{X_3}_{\ell_3 m_3}\hat{a}^{X_4}_{\ell_4 m_4}\>\,.
\end{align}
The inner products are defined
\begin{align}
\<A,\, B\>_k &\equiv  \int_{\curl{V}^T_k} \, \w(k_1,k_2,k_3,k_4) A(k_1,k_2,k_3,k_4) \, B(k_1,k_2,k_3,k_4) d\curl{V}^T_k \,,\\
\<A,\, B\>_l &\equiv \sum_{\ell_i} \w_{\ell_1 \ell_2 \ell_3 \ell_4} \,A_{\ell_1 \ell_2 \ell_3 \ell_4} \, B_{\ell_1 \ell_2 \ell_3 \ell_4}\,,
\end{align}
with the corresponding weights given by
\begin{align}
\bar\w(k_1,k_2,k_3,k_4) = 1\,,\qquad
\w_{\ell_1 \ell_2 \ell_3 \ell_4} = \sum_L \frac{h^2_{\ell_1 \ell_2 L}h^2_{\ell_3 \ell_4 L}}{\(v_{\ell_1}v_{\ell_2}v_{\ell_3}v_{\ell_4}\)^2(2L+1)}\,.
\end{align}
Now the $\ell$-space weight function would naturally scale as $\ell^2$ so in the case of the trispectrum we choose our $v_\ell=(2\ell+1)^{1/4}$. The Fisher matrix scales as $1/\ell^2$ which is the same as the scaling of ${S^T}^2 (k_1,k_2,k_3,k_4)$ so the primordial weight function is $1$. As the primordial shape function now scales as $1/k$ the building blocks of our primordial basis functions should have a $1/k^{1/4}$ scaling to match. However, depending on the particular shape being decomposed, a better choice may be to define
\begin{align}
S(k_1,k_2,k_3,k_4) \equiv \(k_1 k_2 k_3 k_4\)^{9/4} T(k_1,k_2,k_3,k_4)\,,
\end{align}
with weight
\begin{align}
\bar\w(k_1,k_2,k_3,k_4) = \frac{1}{(k_1+k_2+k_3+k_4)^2}\,,
\end{align}
so the primordial shape function is scale invariant which allows primordial basis functions which are scale invariant (eg trigonometric functions or Legendre polynomials) which may improve convergence and the stability of the Cholesky decomposition used to orthogonalise the basis. We need a separable form for the $\ell$-space weight function which is analogous to that for the bispectrum
\begin{align}
\sum_L \frac{h^2_{\ell_1 \ell_2 L}h^2_{\ell_3 \ell_4 L}}{(2L+1)} = \frac{(2\ell_1+1)(2\ell_2+1)(2\ell_3+1)(2\ell_4+1)}{2\(4\pi\)^2} \int d\u P_{\ell_1}(\u)\, P_{\ell_2}(\u)\, P_{\ell_3}(\u)\, P_{\ell_4}(\u)\,.
\end{align}
Now all formula from the bispectrum case translate directly so
\begin{align}
\Gamma^{X_1X_2X_3X_4}_{mn} = \frac{1}{768\pi^2} {\gamma^{X_1X_2X_3X_4}_{mn'}}^{-1} \int d\mu \int x^2 dx &{} \(P^{X_1}_{ii'}(x,\mu)P^{X_2}_{jj'}(x,\mu)P^{X_3}_{kk'}(x,\mu)P^{X_4}_{ll'}(x,\mu) + 23\,\mbox{perms}\)\,,
\end{align}
and
\begin{align}
\hat{\beta}^{X_1X_2X_3X_4}_{n} &= \int d\Omega_n M^{X_1}_{i}(\hat{n}) M^{X_2}_{j}(\hat{n})M^{X_3}_{k}(\hat{n})M^{X_4}_{l}(\hat{n}) - \({\bold M}^{X_1X_2}_{ij}(\hat{n}) M^{X_3}_{k}(\hat{n}) M^{X_4}_{l}(\hat{n}) + 5\,\mbox{perms}\) + \({\bold M}^{X_1 X_2}_{ij}(\hat{n}){\bold M}^{X_3 X_4}_{kl}(\hat{n}) + 2\,\mbox{perms}\)\,.
\end{align}
The estimator becomes
\begin{align}
\curl{E} &= \frac{\sum_{X_i,n} \alpha^{RX_1X_2X_3X_4}_n \beta^{RX_1X_2X_3X_4}_n}{\sum_{X_i,n} {\alpha^{RX_1X_2X_3X_4}_n}^2}\,.
\end{align}
We note that $P^{X}_{ii'}(x,\mu)$, $M^{X}_{i}(\hat{n})$ and $M^{X_1 X_2}_{ij}(\hat{n})$ are those defined by Eq.$\,$(\ref{eq:qtildenew}) and Eq.$\,$(\ref{eq:mapsnew}) but calculated with the new weight, $v_\ell=(2\ell+1)^{1/4}$ rather than $v_\ell=(2\ell+1)^{1/6}$. The multiple trispectrum bases are defined
\begin{align}
Q^{XXXX}_{n\,\ell_1 \ell_2 \ell_3} &= \frac{1}{24} \( q^{X}_{i\ell_1}q^{X}_{j\ell_2}q^{Y}_{k\ell_3}q^{X}_{l\ell_4} + 23 \mbox{perms.}\)\,,\\
Q^{XXXY}_{n\,\ell_1 \ell_2 \ell_3} &= \frac{1}{6}  \( q^{X}_{i\ell_1}q^{X}_{j\ell_2}q^{Y}_{k\ell_3}q^{Y}_{l\ell_4} + 5 \mbox{perms.}\)\,,\\
Q^{XXYY}_{n\,\ell_1 \ell_2 \ell_3} &= \frac{1}{4}  \( q^{X}_{i\ell_1}q^{X}_{j\ell_2}q^{Y}_{k\ell_3}q^{Y}_{l\ell_4} + q^{X}_{i\ell_1}q^{X}_{j\ell_2}q^{Y}_{l\ell_3}q^{Y}_{k\ell_4} + q^{X}_{j\ell_1}q^{X}_{i\ell_2}q^{Y}_{k\ell_3}q^{Y}_{l\ell_4} + q^{X}_{j\ell_1}q^{X}_{i\ell_2}q^{Y}_{l\ell_3}q^{Y}_{k\ell_4}\)\,,\\
Q^{XYYY}_{n\,\ell_1 \ell_2 \ell_3} &= \frac{1}{6}  \( q^{X}_{i\ell_1}q^{Y}_{j\ell_2}q^{Y}_{k\ell_3}q^{Y}_{l\ell_4} + 5 \mbox{perms.} \)\,,
\end{align}
and the orderings now have the more complicated forms
\begin{align}
\begin{array}{c|c|c|c}
Q_n^{XXXX} & Q_n^{XXXY} & Q_n^{XXYY} & Q_n^{XYYY}\\
\hline
\begin{array}{lcllll}
n & \rightarrow & i & j & k & l\\
\hline
0 & \rightarrow & 0 & 0 & 0 & 0\\
1 & \rightarrow & 0 & 0 & 0 & 1\\
2 & \rightarrow & 0 & 0 & 1 & 1\\
3 & \rightarrow & 0 & 1 & 1 & 1\\
4 & \rightarrow & 1 & 1 & 1 & 1\\
5 & \rightarrow & 0 & 0 & 0 & 2\\
\cdots
\end{array}
&
\begin{array}{lcllll}
n & \rightarrow & i & j & k & l\\
\hline
0 & \rightarrow & 0 & 0 & 0 & 0\\
1 & \rightarrow & 0 & 0 & 1 & 0\\
2 & \rightarrow & 0 & 0 & 0 & 1\\
3 & \rightarrow & 0 & 1 & 1 & 0\\
4 & \rightarrow & 0 & 0 & 1 & 1\\
5 & \rightarrow & 1 & 1 & 1 & 0\\
\cdots
\end{array}
&
\begin{array}{lcllll}
n & \rightarrow & i & j & k & l\\
\hline
0 & \rightarrow & 0 & 0 & 0 & 0\\
1 & \rightarrow & 0 & 0 & 0 & 1\\
2 & \rightarrow & 0 & 1 & 0 & 0\\
3 & \rightarrow & 0 & 0 & 1 & 1\\
4 & \rightarrow & 1 & 1 & 0 & 0\\
5 & \rightarrow & 0 & 1 & 0 & 1\\
\cdots
\end{array}
&
\begin{array}{lcllll}
n & \rightarrow & i & j & k & l\\
\hline
0 & \rightarrow & 0 & 0 & 0 & 0\\
1 & \rightarrow & 0 & 0 & 0 & 1\\
2 & \rightarrow & 1 & 0 & 0 & 0\\
3 & \rightarrow & 0 & 0 & 1 & 1\\
4 & \rightarrow & 1 & 0 & 0 & 1\\
5 & \rightarrow & 0 & 1 & 1 & 1\\
\cdots
\end{array}
\end{array}\,.
\end{align}
Finally, we note that the orthogonalisation order for the trispectrum is
\begin{align}\label{eq:triorder}
TTTT &\rightarrow TTTE \rightarrow TTEE \rightarrow TEEE \rightarrow EEEE\,.
\end{align}
So what would have been a $16\times16$ matrix operation containing 22 unique terms is reduced to just 5. It is important to note that using this method, excepting the initial decomposition of the primordial correlators, the diagonal-free trispectrum is no more numerically demanding than the bispectrum as the dimension of all the calculations are the same. The full five dimensional case introduces one extra dimension to the integrals

Simulations can be created by introducing non-Gaussianity to a Gaussian simulation via a cubic term analogous to the quadratic one used for the bispectrum following the temperature only method for the trispectrum presented in \cite{1004.2915} ,
\begin{align}
\hat{a}^T_{\ell m} &= \hat{g}^T_{\ell m} + \frac{1}{24}\sum_{\ell_1 m_1 \ell_2 m_2 \ell_3 m_3} \frac{\int Y_{\ell_1 m_1} Y_{\ell_2 m_2} Y_{\ell_3 m_3} Y_{\ell m}}{v_{\ell_1} v_{\ell_2} v_{\ell_3}  v_\ell}\,f_{NL} \(S^{TTTT}_{\ell_1 \ell_2 \ell_3 \ell}\,  \hat{g}^T_{\ell_1 m_1} \hat{g}^T_{\ell_2 m_2} \hat{g}^T_{\ell_3 m_3}\)\\
\nonumber \hat{a}^E_{\ell m} &= \hat{g}^E_{\ell m} + \frac{1}{24}\sum_{\ell_1 m_1 \ell_2 m_2 \ell_3 m_3} \frac{\int Y_{\ell_1 m_1} Y_{\ell_2 m_2} Y_{\ell_3 m_3} Y_{\ell m}}{v_{\ell_1} v_{\ell_2} v_{\ell_3}  v_\ell}\, f_{NL}\(4 S^{TTTE}_{\ell_1 \ell_2 \ell_3 \ell}\,  \hat{g}^T_{\ell_1 m_1} \hat{g}^T_{\ell_2 m_2}  \hat{g}^T_{\ell_3 m_3}+ 6 S^{TTEE}_{\ell_1 \ell_2 \ell_3 \ell}\,  \hat{g}^T_{\ell_1 m_1} \hat{g}^T_{\ell_2 m_2} \hat{g}^E_{\ell_3 m_3}\right.\\
 &\quad\quad\quad\quad\quad\quad\quad\quad\quad\quad\quad\quad\quad\quad\quad\quad\quad\quad\quad\quad+ \left.4 S^{TEEE}_{\ell_1 \ell_2 \ell_3 \ell}\,  \hat{g}^T_{\ell_1 m_1} \hat{g}^E_{\ell_2 m_2}  \hat{g}^E_{\ell_3 m_3} + S^{EEEE}_{\ell_1 \ell_2 \ell_3 \ell}\,  \hat{g}^E_{\ell_1 m_1} \hat{g}^E_{\ell_2 m_2}  \hat{g}^E_{\ell_3 m_3} \)\,.
\end{align}
Which, with the analogous definitions,
\begin{align}
G^X_{i}(\hat{n}) \equiv \sum_{lm} \frac{1}{v_{\ell}} q^X_{i\ell} \hat{g}^X_{\ell m} Y_{\ell m}(\hat{n})\,, \quad {\bold G}^{X_1X_2X_3}_{\ell m\,|\, ijk} = \frac{1}{v_{\ell}}\int Y_{\ell m}(\un) G^{X_1}_{i}(\un) G^{X_2}_{j}(\un)G^{X_3}_{k}(\un) \, d^2\un \,,
\end{align}
becomes after substitution of the separable form for the shape functions,
\begin{align}
\hat{a}^T_{\ell m} &= \hat{g}^T_{\ell m} + f_{NL} \sum_{n \rightarrow ijks} \frac{1}{96} \a^{TTTT}_n\( {\bold G}^{TTT}_{ijk} q^T_{s} + {\bold G}^{TTT}_{jks} q^T_{i} + {\bold G}^{TTT}_{ksi} q^T_{j} + {\bold G}^{TTT}_{sij} q^T_{k}  \)_{\ell m} \\
\hat{a}^E_{\ell m} &= \hat{g}^E_{\ell m} + f_{NL} \sum_{n \rightarrow ijk} \(\frac{1}{6}\a^{TTTE}_n {\bold G}^{TTT}_{ijk} q^E_{s} + \frac{1}{8}\a^{TTEE}_n \({\bold G}^{TTE}_{ijk} q^E_{s} + {\bold G}^{TTE}_{ijs} q^E_{k}\) + \frac{1}{18}\a^{TEEE}_n \({\bold G}^{TEE}_{ijk} q^E_{s} + {\bold G}^{TEE}_{iks} q^E_{j} + {\bold G}^{TEE}_{isj} q^E_{k}\)\right.\\
&\quad\quad\quad\quad\quad\quad\quad\quad+ \left. \frac{1}{96}\a^{EEEE}_n \(  {\bold G}^{EEE}_{ijk} q^E_{s} + {\bold G}^{EEE}_{jks} q^E_{i} + {\bold G}^{EEE}_{ksi} q^E_{j} + {\bold G}^{EEE}_{sij} q^E_{k} \) \)_{\ell m}\,.
\end{align}
It should be noted that if both bispectra and trispectra wish to be simulated in the same map then we need to subtract off the bispectrum squared contribution from the trispectrum part. 
\bibliographystyle{unsrt}
\bibliography{Polarisation}

\end{document}

%% file: Planck.tex
\def\setsymbol#1#2{\expandafter\def\csname #1\endcsname{#2}}
\def\getsymbol#1{\csname #1\endcsname}

\newbox\tablebox    \newdimen\tablewidth
\def\leaderfil{\leaders\hbox to 5pt{\hss.\hss}\hfil}
\def\tablenote#1 #2\par{\begingroup \parindent=0.8em
    \abovedisplayshortskip=0pt\belowdisplayshortskip=0pt
    \noindent
    $$\hss\vbox{\hsize\tablewidth \hangindent=\parindent \hangafter=1 \noindent
    \hbox to \parindent{$^#1$\hss}\strut#2\strut\par}\hss$$
    \endgroup}

\def\L2{\ifmmode L_2\else $L_2$\fi}

\def\DeltaT{\ifmmode \Delta T\else $\Delta T$\fi}
\def\deltat{\ifmmode \Delta t\else $\Delta t$\fi}
\def\fknee{\ifmmode f_{\rm knee}\else $f_{\rm knee}$\fi}
\def\Fmax{\ifmmode F_{\rm max}\else $F_{\rm max}$\fi}
\def\solar{\ifmmode{\rm M}_{\mathord\odot}\else${\rm M}_{\mathord\odot}$\fi}
\def\Msolar{\ifmmode{\rm M}_{\mathord\odot}\else${\rm M}_{\mathord\odot}$\fi}
\def\Lsolar{\ifmmode{\rm L}_{\mathord\odot}\else${\rm L}_{\mathord\odot}$\fi}

\def\inv{\ifmmode^{-1}\else$^{-1}$\fi}
\def\mo{\ifmmode^{-1}\else$^{-1}$\fi}
\def\sup#1{\ifmmode ^{\rm #1}\else $^{\rm #1}$\fi}
\def\expo#1{\ifmmode \times 10^{#1}\else $\times 10^{#1}$\fi}
\def\,{\thinspace}
\def\lsim{\mathrel{\raise .4ex\hbox{\rlap{$<$}\lower 1.2ex\hbox{$\sim$}}}}
\def\gsim{\mathrel{\raise .4ex\hbox{\rlap{$>$}\lower 1.2ex\hbox{$\sim$}}}}

\def\simprop{\mathrel{\raise .4ex\hbox{\rlap{$\propto$}\lower 1.2ex\hbox{$\sim$}}}}
\def\deg{\ifmmode^\circ\else$^\circ$\fi}
\def\pdeg{\ifmmode $\setbox0=\hbox{$^{\circ}$}\rlap{\hskip.11\wd0 .}$^{\circ}
          \else \setbox0=\hbox{$^{\circ}$}\rlap{\hskip.11\wd0 .}$^{\circ}$\fi}
\def\arcs{\ifmmode {^{\scriptstyle\prime\prime}}
          \else $^{\scriptstyle\prime\prime}$\fi}
\def\arcm{\ifmmode {^{\scriptstyle\prime}}
          \else $^{\scriptstyle\prime}$\fi}
\newdimen\sa  \newdimen\sb
\def\parcs{\sa=.07em \sb=.03em
     \ifmmode \hbox{\rlap{.}}^{\scriptstyle\prime\kern -\sb\prime}\hbox{\kern -\sa}
     \else \rlap{.}$^{\scriptstyle\prime\kern -\sb\prime}$\kern -\sa\fi}
\def\parcm{\sa=.08em \sb=.03em
     \ifmmode \hbox{\rlap{.}\kern\sa}^{\scriptstyle\prime}\hbox{\kern-\sb}
     \else \rlap{.}\kern\sa$^{\scriptstyle\prime}$\kern-\sb\fi}
\def\ra[#1 #2 #3.#4]{#1\sup{h}#2\sup{m}#3\sup{s}\llap.#4}
\def\dec[#1 #2 #3.#4]{#1\deg#2\arcm#3\arcs\llap.#4}
\def\deco[#1 #2 #3]{#1\deg#2\arcm#3\arcs}
\def\rra[#1 #2]{#1\sup{h}#2\sup{m}}

\def\dots{\relax\ifmmode \ldots\else $\ldots$\fi}
\def\WHzsr{\ifmmode $W\,Hz\mo\,sr\mo$\else W\,Hz\mo\,sr\mo\fi}
\def\mHz{\ifmmode $\,mHz$\else \,mHz\fi}
\def\GHz{\ifmmode $\,GHz$\else \,GHz\fi}
\def\mKs{\ifmmode $\,mK\,s$^{1/2}\else \,mK\,s$^{1/2}$\fi}
\def\muKs{\ifmmode \,\mu$K\,s$^{1/2}\else \,$\mu$K\,s$^{1/2}$\fi}
\def\muKRJs{\ifmmode \,\mu$K$_{\rm RJ}$\,s$^{1/2}\else \,$\mu$K$_{\rm RJ}$\,s$^{1/2}$\fi}
\def\muKHz{\ifmmode \,\mu$K\,Hz$^{-1/2}\else \,$\mu$K\,Hz$^{-1/2}$\fi}
\def\MJysr{\ifmmode \,$MJy\,sr\mo$\else \,MJy\,sr\mo\fi}
\def\MJysrmK{\ifmmode \,$MJy\,sr\mo$\,mK$_{\rm CMB}\mo\else \,MJy\,sr\mo\,mK$_{\rm CMB}\mo$\fi}
\def\microns{\ifmmode \,\mu$m$\else \,$\mu$m\fi}

\def\muK{\ifmmode \,\mu$K$\else \,$\mu$\hbox{K}\fi}
\def\microK{\ifmmode \,\mu$K$\else \,$\mu$\hbox{K}\fi}
\def\muW{\ifmmode \,\mu$W$\else \,$\mu$\hbox{W}\fi}
\def\kms{\ifmmode $\,km\,s$^{-1}\else \,km\,s$^{-1}$\fi}
\def\kmsMpc{\ifmmode $\,\kms\,Mpc\mo$\else \,\kms\,Mpc\mo\fi}
\setsymbol{LFI:center:frequency:70GHz:units}{70.3\,GHz}
\setsymbol{LFI:center:frequency:44GHz:units}{44.1\,GHz}
\setsymbol{LFI:center:frequency:30GHz:units}{28.5\,GHz}

\setsymbol{LFI:center:frequency:70GHz}{70.3}
\setsymbol{LFI:center:frequency:44GHz}{44.1}
\setsymbol{LFI:center:frequency:30GHz}{28.5}

\setsymbol{LFI:center:frequency:LFI18:Rad:M:units}{71.7\GHz}
\setsymbol{LFI:center:frequency:LFI19:Rad:M:units}{67.5\GHz}
\setsymbol{LFI:center:frequency:LFI20:Rad:M:units}{69.2\GHz}
\setsymbol{LFI:center:frequency:LFI21:Rad:M:units}{70.4\GHz}
\setsymbol{LFI:center:frequency:LFI22:Rad:M:units}{71.5\GHz}
\setsymbol{LFI:center:frequency:LFI23:Rad:M:units}{70.8\GHz}
\setsymbol{LFI:center:frequency:LFI24:Rad:M:units}{44.4\GHz}
\setsymbol{LFI:center:frequency:LFI25:Rad:M:units}{44.0\GHz}
\setsymbol{LFI:center:frequency:LFI26:Rad:M:units}{43.9\GHz}
\setsymbol{LFI:center:frequency:LFI27:Rad:M:units}{28.3\GHz}
\setsymbol{LFI:center:frequency:LFI28:Rad:M:units}{28.8\GHz}
\setsymbol{LFI:center:frequency:LFI18:Rad:S:units}{70.1\GHz}
\setsymbol{LFI:center:frequency:LFI19:Rad:S:units}{69.6\GHz}
\setsymbol{LFI:center:frequency:LFI20:Rad:S:units}{69.5\GHz}
\setsymbol{LFI:center:frequency:LFI21:Rad:S:units}{69.5\GHz}
\setsymbol{LFI:center:frequency:LFI22:Rad:S:units}{72.8\GHz}
\setsymbol{LFI:center:frequency:LFI23:Rad:S:units}{71.3\GHz}
\setsymbol{LFI:center:frequency:LFI24:Rad:S:units}{44.1\GHz}
\setsymbol{LFI:center:frequency:LFI25:Rad:S:units}{44.1\GHz}
\setsymbol{LFI:center:frequency:LFI26:Rad:S:units}{44.1\GHz}
\setsymbol{LFI:center:frequency:LFI27:Rad:S:units}{28.5\GHz}
\setsymbol{LFI:center:frequency:LFI28:Rad:S:units}{28.2\GHz}

\setsymbol{LFI:center:frequency:LFI18:Rad:M}{71.7}
\setsymbol{LFI:center:frequency:LFI19:Rad:M}{67.5}
\setsymbol{LFI:center:frequency:LFI20:Rad:M}{69.2}
\setsymbol{LFI:center:frequency:LFI21:Rad:M}{70.4}
\setsymbol{LFI:center:frequency:LFI22:Rad:M}{71.5}
\setsymbol{LFI:center:frequency:LFI23:Rad:M}{70.8}
\setsymbol{LFI:center:frequency:LFI24:Rad:M}{44.4}
\setsymbol{LFI:center:frequency:LFI25:Rad:M}{44.0}
\setsymbol{LFI:center:frequency:LFI26:Rad:M}{43.9}
\setsymbol{LFI:center:frequency:LFI27:Rad:M}{28.3}
\setsymbol{LFI:center:frequency:LFI28:Rad:M}{28.8}
\setsymbol{LFI:center:frequency:LFI18:Rad:S}{70.1}
\setsymbol{LFI:center:frequency:LFI19:Rad:S}{69.6}
\setsymbol{LFI:center:frequency:LFI20:Rad:S}{69.5}
\setsymbol{LFI:center:frequency:LFI21:Rad:S}{69.5}
\setsymbol{LFI:center:frequency:LFI22:Rad:S}{72.8}
\setsymbol{LFI:center:frequency:LFI23:Rad:S}{71.3}
\setsymbol{LFI:center:frequency:LFI24:Rad:S}{44.1}
\setsymbol{LFI:center:frequency:LFI25:Rad:S}{44.1}
\setsymbol{LFI:center:frequency:LFI26:Rad:S}{44.1}
\setsymbol{LFI:center:frequency:LFI27:Rad:S}{28.5}
\setsymbol{LFI:center:frequency:LFI28:Rad:S}{28.2}

\setsymbol{LFI:white:noise:sensitivity:70GHz:units}{134.7\muKs}
\setsymbol{LFI:white:noise:sensitivity:44GHz:units}{164.7\muKs}
\setsymbol{LFI:white:noise:sensitivity:30GHz:units}{143.4\muKs}

\setsymbol{LFI:white:noise:sensitivity:70GHz}{134.7}
\setsymbol{LFI:white:noise:sensitivity:44GHz}{164.7}
\setsymbol{LFI:white:noise:sensitivity:30GHz}{143.4}

\setsymbol{LFI:white:noise:sensitivity:LFI18:Rad:M:units}{512.0\muKs}
\setsymbol{LFI:white:noise:sensitivity:LFI19:Rad:M:units}{581.4\muKs}
\setsymbol{LFI:white:noise:sensitivity:LFI20:Rad:M:units}{590.8\muKs}
\setsymbol{LFI:white:noise:sensitivity:LFI21:Rad:M:units}{455.2\muKs}
\setsymbol{LFI:white:noise:sensitivity:LFI22:Rad:M:units}{492.0\muKs}
\setsymbol{LFI:white:noise:sensitivity:LFI23:Rad:M:units}{507.7\muKs}
\setsymbol{LFI:white:noise:sensitivity:LFI24:Rad:M:units}{462.2\muKs}
\setsymbol{LFI:white:noise:sensitivity:LFI25:Rad:M:units}{413.6\muKs}
\setsymbol{LFI:white:noise:sensitivity:LFI26:Rad:M:units}{478.6\muKs}
\setsymbol{LFI:white:noise:sensitivity:LFI27:Rad:M:units}{277.7\muKs}
\setsymbol{LFI:white:noise:sensitivity:LFI28:Rad:M:units}{312.3\muKs}
\setsymbol{LFI:white:noise:sensitivity:LFI18:Rad:S:units}{465.7\muKs}
\setsymbol{LFI:white:noise:sensitivity:LFI19:Rad:S:units}{555.6\muKs}
\setsymbol{LFI:white:noise:sensitivity:LFI20:Rad:S:units}{623.2\muKs}
\setsymbol{LFI:white:noise:sensitivity:LFI21:Rad:S:units}{564.1\muKs}
\setsymbol{LFI:white:noise:sensitivity:LFI22:Rad:S:units}{534.4\muKs}
\setsymbol{LFI:white:noise:sensitivity:LFI23:Rad:S:units}{542.4\muKs}
\setsymbol{LFI:white:noise:sensitivity:LFI24:Rad:S:units}{399.2\muKs}
\setsymbol{LFI:white:noise:sensitivity:LFI25:Rad:S:units}{392.6\muKs}
\setsymbol{LFI:white:noise:sensitivity:LFI26:Rad:S:units}{418.6\muKs}
\setsymbol{LFI:white:noise:sensitivity:LFI27:Rad:S:units}{302.9\muKs}
\setsymbol{LFI:white:noise:sensitivity:LFI28:Rad:S:units}{285.3\muKs}

\setsymbol{LFI:white:noise:sensitivity:LFI18:Rad:M}{512.0}
\setsymbol{LFI:white:noise:sensitivity:LFI19:Rad:M}{581.4}
\setsymbol{LFI:white:noise:sensitivity:LFI20:Rad:M}{590.8}
\setsymbol{LFI:white:noise:sensitivity:LFI21:Rad:M}{455.2}
\setsymbol{LFI:white:noise:sensitivity:LFI22:Rad:M}{492.0}
\setsymbol{LFI:white:noise:sensitivity:LFI23:Rad:M}{507.7}
\setsymbol{LFI:white:noise:sensitivity:LFI24:Rad:M}{462.2}
\setsymbol{LFI:white:noise:sensitivity:LFI25:Rad:M}{413.6}
\setsymbol{LFI:white:noise:sensitivity:LFI26:Rad:M}{478.6}
\setsymbol{LFI:white:noise:sensitivity:LFI27:Rad:M}{277.7}
\setsymbol{LFI:white:noise:sensitivity:LFI28:Rad:M}{312.3}
\setsymbol{LFI:white:noise:sensitivity:LFI18:Rad:S}{465.7}
\setsymbol{LFI:white:noise:sensitivity:LFI19:Rad:S}{555.6}
\setsymbol{LFI:white:noise:sensitivity:LFI20:Rad:S}{623.2}
\setsymbol{LFI:white:noise:sensitivity:LFI21:Rad:S}{564.1}
\setsymbol{LFI:white:noise:sensitivity:LFI22:Rad:S}{534.4}
\setsymbol{LFI:white:noise:sensitivity:LFI23:Rad:S}{542.4}
\setsymbol{LFI:white:noise:sensitivity:LFI24:Rad:S}{399.2}
\setsymbol{LFI:white:noise:sensitivity:LFI25:Rad:S}{392.6}
\setsymbol{LFI:white:noise:sensitivity:LFI26:Rad:S}{418.6}
\setsymbol{LFI:white:noise:sensitivity:LFI27:Rad:S}{302.9}
\setsymbol{LFI:white:noise:sensitivity:LFI28:Rad:S}{285.3}

\setsymbol{LFI:knee:frequency:70GHz:units}{29.5\mHz}
\setsymbol{LFI:knee:frequency:44GHz:units}{56.2\mHz}
\setsymbol{LFI:knee:frequency:30GHz:units}{113.7\mHz}

\setsymbol{LFI:knee:frequency:70GHz}{29.5}
\setsymbol{LFI:knee:frequency:44GHz}{56.2}
\setsymbol{LFI:knee:frequency:30GHz}{113.7}

\setsymbol{LFI:knee:frequency:LFI18:Rad:M:units}{16.3\mHz}
\setsymbol{LFI:knee:frequency:LFI19:Rad:M:units}{15.1\mHz}
\setsymbol{LFI:knee:frequency:LFI20:Rad:M:units}{18.7\mHz}
\setsymbol{LFI:knee:frequency:LFI21:Rad:M:units}{37.2\mHz}
\setsymbol{LFI:knee:frequency:LFI22:Rad:M:units}{12.7\mHz}
\setsymbol{LFI:knee:frequency:LFI23:Rad:M:units}{34.6\mHz}
\setsymbol{LFI:knee:frequency:LFI24:Rad:M:units}{46.2\mHz}
\setsymbol{LFI:knee:frequency:LFI25:Rad:M:units}{24.9\mHz}
\setsymbol{LFI:knee:frequency:LFI26:Rad:M:units}{67.6\mHz}
\setsymbol{LFI:knee:frequency:LFI27:Rad:M:units}{187.4\mHz}
\setsymbol{LFI:knee:frequency:LFI28:Rad:M:units}{122.2\mHz}
\setsymbol{LFI:knee:frequency:LFI18:Rad:S:units}{17.7\mHz}
\setsymbol{LFI:knee:frequency:LFI19:Rad:S:units}{22.0\mHz}
\setsymbol{LFI:knee:frequency:LFI20:Rad:S:units}{8.7\mHz}
\setsymbol{LFI:knee:frequency:LFI21:Rad:S:units}{25.9\mHz}
\setsymbol{LFI:knee:frequency:LFI22:Rad:S:units}{15.8\mHz}
\setsymbol{LFI:knee:frequency:LFI23:Rad:S:units}{129.8\mHz}
\setsymbol{LFI:knee:frequency:LFI24:Rad:S:units}{100.9\mHz}
\setsymbol{LFI:knee:frequency:LFI25:Rad:S:units}{38.9\mHz}
\setsymbol{LFI:knee:frequency:LFI26:Rad:S:units}{58.9\mHz}
\setsymbol{LFI:knee:frequency:LFI27:Rad:S:units}{104.4\mHz}
\setsymbol{LFI:knee:frequency:LFI28:Rad:S:units}{40.7\mHz}

\setsymbol{LFI:knee:frequency:LFI18:Rad:M}{16.3}
\setsymbol{LFI:knee:frequency:LFI19:Rad:M}{15.1}
\setsymbol{LFI:knee:frequency:LFI20:Rad:M}{18.7}
\setsymbol{LFI:knee:frequency:LFI21:Rad:M}{37.2}
\setsymbol{LFI:knee:frequency:LFI22:Rad:M}{12.7}
\setsymbol{LFI:knee:frequency:LFI23:Rad:M}{34.6}
\setsymbol{LFI:knee:frequency:LFI24:Rad:M}{46.2}
\setsymbol{LFI:knee:frequency:LFI25:Rad:M}{24.9}
\setsymbol{LFI:knee:frequency:LFI26:Rad:M}{67.6}
\setsymbol{LFI:knee:frequency:LFI27:Rad:M}{187.4}
\setsymbol{LFI:knee:frequency:LFI28:Rad:M}{122.2}
\setsymbol{LFI:knee:frequency:LFI18:Rad:S}{17.7}
\setsymbol{LFI:knee:frequency:LFI19:Rad:S}{22.0}
\setsymbol{LFI:knee:frequency:LFI20:Rad:S}{8.7}
\setsymbol{LFI:knee:frequency:LFI21:Rad:S}{25.9}
\setsymbol{LFI:knee:frequency:LFI22:Rad:S}{15.8}
\setsymbol{LFI:knee:frequency:LFI23:Rad:S}{129.8}
\setsymbol{LFI:knee:frequency:LFI24:Rad:S}{100.9}
\setsymbol{LFI:knee:frequency:LFI25:Rad:S}{38.9}
\setsymbol{LFI:knee:frequency:LFI26:Rad:S}{58.9}
\setsymbol{LFI:knee:frequency:LFI27:Rad:S}{104.4}
\setsymbol{LFI:knee:frequency:LFI28:Rad:S}{40.7}

\setsymbol{LFI:slope:70GHz:units}{$-1.03$\mHz}
\setsymbol{LFI:slope:44GHz:units}{$-0.89$\mHz}
\setsymbol{LFI:slope:30GHz:units}{$-0.87$\mHz}

\setsymbol{LFI:slope:70GHz}{$-1.03$}
\setsymbol{LFI:slope:44GHz}{$-0.89$}
\setsymbol{LFI:slope:30GHz}{$-0.87$}

\setsymbol{LFI:slope:LFI18:Rad:M:units}{$-1.04$\mHz}
\setsymbol{LFI:slope:LFI19:Rad:M:units}{$-1.09$\mHz}
\setsymbol{LFI:slope:LFI20:Rad:M:units}{$-0.69$\mHz}
\setsymbol{LFI:slope:LFI21:Rad:M:units}{$-1.56$\mHz}
\setsymbol{LFI:slope:LFI22:Rad:M:units}{$-1.01$\mHz}
\setsymbol{LFI:slope:LFI23:Rad:M:units}{$-0.96$\mHz}
\setsymbol{LFI:slope:LFI24:Rad:M:units}{$-0.83$\mHz}
\setsymbol{LFI:slope:LFI25:Rad:M:units}{$-0.91$\mHz}
\setsymbol{LFI:slope:LFI26:Rad:M:units}{$-0.95$\mHz}
\setsymbol{LFI:slope:LFI27:Rad:M:units}{$-0.87$\mHz}
\setsymbol{LFI:slope:LFI28:Rad:M:units}{$-0.88$\mHz}
\setsymbol{LFI:slope:LFI18:Rad:S:units}{$-1.15$\mHz}
\setsymbol{LFI:slope:LFI19:Rad:S:units}{$-1.00$\mHz}
\setsymbol{LFI:slope:LFI20:Rad:S:units}{$-0.95$\mHz}
\setsymbol{LFI:slope:LFI21:Rad:S:units}{$-0.92$\mHz}
\setsymbol{LFI:slope:LFI22:Rad:S:units}{$-1.01$\mHz}
\setsymbol{LFI:slope:LFI23:Rad:S:units}{$-0.95$\mHz}
\setsymbol{LFI:slope:LFI24:Rad:S:units}{$-0.73$\mHz}
\setsymbol{LFI:slope:LFI25:Rad:S:units}{$-1.16$\mHz}
\setsymbol{LFI:slope:LFI26:Rad:S:units}{$-0.79$\mHz}
\setsymbol{LFI:slope:LFI27:Rad:S:units}{$-0.82$\mHz}
\setsymbol{LFI:slope:LFI28:Rad:S:units}{$-0.91$\mHz}

\setsymbol{LFI:slope:LFI18:Rad:M}{$-1.04$}
\setsymbol{LFI:slope:LFI19:Rad:M}{$-1.09$}
\setsymbol{LFI:slope:LFI20:Rad:M}{$-0.69$}
\setsymbol{LFI:slope:LFI21:Rad:M}{$-1.56$}
\setsymbol{LFI:slope:LFI22:Rad:M}{$-1.01$}
\setsymbol{LFI:slope:LFI23:Rad:M}{$-0.96$}
\setsymbol{LFI:slope:LFI24:Rad:M}{$-0.83$}
\setsymbol{LFI:slope:LFI25:Rad:M}{$-0.91$}
\setsymbol{LFI:slope:LFI26:Rad:M}{$-0.95$}
\setsymbol{LFI:slope:LFI27:Rad:M}{$-0.87$}
\setsymbol{LFI:slope:LFI28:Rad:M}{$-0.88$}
\setsymbol{LFI:slope:LFI18:Rad:S}{$-1.15$}
\setsymbol{LFI:slope:LFI19:Rad:S}{$-1.00$}
\setsymbol{LFI:slope:LFI20:Rad:S}{$-0.95$}
\setsymbol{LFI:slope:LFI21:Rad:S}{$-0.92$}
\setsymbol{LFI:slope:LFI22:Rad:S}{$-1.01$}
\setsymbol{LFI:slope:LFI23:Rad:S}{$-0.95$}
\setsymbol{LFI:slope:LFI24:Rad:S}{$-0.73$}
\setsymbol{LFI:slope:LFI25:Rad:S}{$-1.16$}
\setsymbol{LFI:slope:LFI26:Rad:S}{$-0.79$}
\setsymbol{LFI:slope:LFI27:Rad:S}{$-0.82$}
\setsymbol{LFI:slope:LFI28:Rad:S}{$-0.91$}

\setsymbol{LFI:FWHM:70GHz:units}{13\parcm01}
\setsymbol{LFI:FWHM:44GHz:units}{27\parcm92}
\setsymbol{LFI:FWHM:30GHz:units}{32\parcm65}

\setsymbol{LFI:FWHM:70GHz}{13.01}
\setsymbol{LFI:FWHM:44GHz}{27.92}
\setsymbol{LFI:FWHM:30GHz}{32.65}

\setsymbol{LFI:FWHM:LFI18:units}{13\parcm39}
\setsymbol{LFI:FWHM:LFI19:units}{13\parcm01}
\setsymbol{LFI:FWHM:LFI20:units}{12\parcm75}
\setsymbol{LFI:FWHM:LFI21:units}{12\parcm74}
\setsymbol{LFI:FWHM:LFI22:units}{12\parcm87}
\setsymbol{LFI:FWHM:LFI23:units}{13\parcm27}
\setsymbol{LFI:FWHM:LFI24:units}{22\parcm98}
\setsymbol{LFI:FWHM:LFI25:units}{30\parcm46}
\setsymbol{LFI:FWHM:LFI26:units}{30\parcm31}
\setsymbol{LFI:FWHM:LFI27:units}{32\parcm65}
\setsymbol{LFI:FWHM:LFI28:units}{32\parcm66}

\setsymbol{LFI:FWHM:LFI18}{13.39}
\setsymbol{LFI:FWHM:LFI19}{13.01}
\setsymbol{LFI:FWHM:LFI20}{12.75}
\setsymbol{LFI:FWHM:LFI21}{12.74}
\setsymbol{LFI:FWHM:LFI22}{12.87}
\setsymbol{LFI:FWHM:LFI23}{13.27}
\setsymbol{LFI:FWHM:LFI24}{22.98}
\setsymbol{LFI:FWHM:LFI25}{30.46}
\setsymbol{LFI:FWHM:LFI26}{30.31}
\setsymbol{LFI:FWHM:LFI27}{32.65}
\setsymbol{LFI:FWHM:LFI28}{32.66}

\setsymbol{LFI:FWHM:uncertainty:LFI18:units}{0.170\arcm}
\setsymbol{LFI:FWHM:uncertainty:LFI19:units}{0.174\arcm}
\setsymbol{LFI:FWHM:uncertainty:LFI20:units}{0.170\arcm}
\setsymbol{LFI:FWHM:uncertainty:LFI21:units}{0.156\arcm}
\setsymbol{LFI:FWHM:uncertainty:LFI22:units}{0.164\arcm}
\setsymbol{LFI:FWHM:uncertainty:LFI23:units}{0.171\arcm}
\setsymbol{LFI:FWHM:uncertainty:LFI24:units}{0.652\arcm}
\setsymbol{LFI:FWHM:uncertainty:LFI25:units}{1.075\arcm}
\setsymbol{LFI:FWHM:uncertainty:LFI26:units}{1.131\arcm}
\setsymbol{LFI:FWHM:uncertainty:LFI27:units}{1.266\arcm}
\setsymbol{LFI:FWHM:uncertainty:LFI28:units}{1.287\arcm}

\setsymbol{LFI:FWHM:uncertainty:LFI18}{0.170}
\setsymbol{LFI:FWHM:uncertainty:LFI19}{0.174}
\setsymbol{LFI:FWHM:uncertainty:LFI20}{0.170}
\setsymbol{LFI:FWHM:uncertainty:LFI21}{0.156}
\setsymbol{LFI:FWHM:uncertainty:LFI22}{0.164}
\setsymbol{LFI:FWHM:uncertainty:LFI23}{0.171}
\setsymbol{LFI:FWHM:uncertainty:LFI24}{0.652}
\setsymbol{LFI:FWHM:uncertainty:LFI25}{1.075}
\setsymbol{LFI:FWHM:uncertainty:LFI26}{1.131}
\setsymbol{LFI:FWHM:uncertainty:LFI27}{1.266}
\setsymbol{LFI:FWHM:uncertainty:LFI28}{1.287}

\setsymbol{HFI:center:frequency:100GHz:units}{100\,GHz}
\setsymbol{HFI:center:frequency:143GHz:units}{143\,GHz}
\setsymbol{HFI:center:frequency:217GHz:units}{217\,GHz}
\setsymbol{HFI:center:frequency:353GHz:units}{353\,GHz}
\setsymbol{HFI:center:frequency:545GHz:units}{545\,GHz}
\setsymbol{HFI:center:frequency:857GHz:units}{857\,GHz}

\setsymbol{HFI:center:frequency:100GHz}{100}
\setsymbol{HFI:center:frequency:143GHz}{143}
\setsymbol{HFI:center:frequency:217GHz}{217}
\setsymbol{HFI:center:frequency:353GHz}{353}
\setsymbol{HFI:center:frequency:545GHz}{545}
\setsymbol{HFI:center:frequency:857GHz}{857}

\setsymbol{HFI:Ndetectors:100GHz}{8}
\setsymbol{HFI:Ndetectors:143GHz}{11}
\setsymbol{HFI:Ndetectors:217GHz}{12}
\setsymbol{HFI:Ndetectors:353GHz}{12}
\setsymbol{HFI:Ndetectors:545GHz}{3}
\setsymbol{HFI:Ndetectors:857GHz}{4}

\setsymbol{HFI:FWHM:Maps:100GHz:units}{9\parcm88}
\setsymbol{HFI:FWHM:Maps:143GHz:units}{7\parcm18}
\setsymbol{HFI:FWHM:Maps:217GHz:units}{4\parcm87}
\setsymbol{HFI:FWHM:Maps:353GHz:units}{4\parcm65}
\setsymbol{HFI:FWHM:Maps:545GHz:units}{4\parcm72}
\setsymbol{HFI:FWHM:Maps:857GHz:units}{4\parcm39}
\setsymbol{HFI:FWHM:Maps:100GHz}{9.88}
\setsymbol{HFI:FWHM:Maps:143GHz}{7.18}
\setsymbol{HFI:FWHM:Maps:217GHz}{4.87}
\setsymbol{HFI:FWHM:Maps:353GHz}{4.65}
\setsymbol{HFI:FWHM:Maps:545GHz}{4.72}
\setsymbol{HFI:FWHM:Maps:857GHz}{4.39}

\setsymbol{HFI:beam:ellipticity:Maps:100GHz}{1.15}
\setsymbol{HFI:beam:ellipticity:Maps:143GHz}{1.01}
\setsymbol{HFI:beam:ellipticity:Maps:217GHz}{1.06}
\setsymbol{HFI:beam:ellipticity:Maps:353GHz}{1.05}
\setsymbol{HFI:beam:ellipticity:Maps:545GHz}{1.14}
\setsymbol{HFI:beam:ellipticity:Maps:857GHz}{1.19}

\setsymbol{HFI:FWHM:Mars:100GHz:units}{9\parcm37}
\setsymbol{HFI:FWHM:Mars:143GHz:units}{7\parcm04}
\setsymbol{HFI:FWHM:Mars:217GHz:units}{4\parcm68}
\setsymbol{HFI:FWHM:Mars:353GHz:units}{4\parcm43}
\setsymbol{HFI:FWHM:Mars:545GHz:units}{3\parcm80}
\setsymbol{HFI:FWHM:Mars:857GHz:units}{3\parcm67}

\setsymbol{HFI:FWHM:Mars:100GHz}{9.37}
\setsymbol{HFI:FWHM:Mars:143GHz}{7.04}
\setsymbol{HFI:FWHM:Mars:217GHz}{4.68}
\setsymbol{HFI:FWHM:Mars:353GHz}{4.43}
\setsymbol{HFI:FWHM:Mars:545GHz}{3.80}
\setsymbol{HFI:FWHM:Mars:857GHz}{3.67}

\setsymbol{HFI:beam:ellipticity:Mars:100GHz}{1.18}
\setsymbol{HFI:beam:ellipticity:Mars:143GHz}{1.03}
\setsymbol{HFI:beam:ellipticity:Mars:217GHz}{1.14}
\setsymbol{HFI:beam:ellipticity:Mars:353GHz}{1.09}
\setsymbol{HFI:beam:ellipticity:Mars:545GHz}{1.25}
\setsymbol{HFI:beam:ellipticity:Mars:857GHz}{1.03}

\setsymbol{HFI:CMB:relative:calibration:100GHz}{$\lsim 1\%$}
\setsymbol{HFI:CMB:relative:calibration:143GHz}{$\lsim 1\%$}
\setsymbol{HFI:CMB:relative:calibration:217GHz}{$\lsim 1\%$}
\setsymbol{HFI:CMB:relative:calibration:353GHz}{$\lsim 1\%$}
\setsymbol{HFI:CMB:relative:calibration:545GHz}{}
\setsymbol{HFI:CMB:relative:calibration:857GHz}{}

\setsymbol{HFI:CMB:absolute:calibration:100GHz}{$\lsim 2\%$}
\setsymbol{HFI:CMB:absolute:calibration:143GHz}{$\lsim 2\%$}
\setsymbol{HFI:CMB:absolute:calibration:217GHz}{$\lsim 2\%$}
\setsymbol{HFI:CMB:absolute:calibration:353GHz}{$\lsim 2\%$}
\setsymbol{HFI:CMB:absolute:calibration:545GHz}{}
\setsymbol{HFI:CMB:absolute:calibration:857GHz}{}

\setsymbol{HFI:FIRAS:gain:calibration:accuracy:statistical:100GHz}{}
\setsymbol{HFI:FIRAS:gain:calibration:accuracy:statistical:143GHz}{}
\setsymbol{HFI:FIRAS:gain:calibration:accuracy:statistical:217GHz}{}
\setsymbol{HFI:FIRAS:gain:calibration:accuracy:statistical:353GHz}{2.5\%}
\setsymbol{HFI:FIRAS:gain:calibration:accuracy:statistical:545GHz}{1\%}
\setsymbol{HFI:FIRAS:gain:calibration:accuracy:statistical:857GHz}{0.5\%}

\setsymbol{HFI:FIRAS:gain:calibration:accuracy:systematic:100GHz}{}
\setsymbol{HFI:FIRAS:gain:calibration:accuracy:systematic:143GHz}{}
\setsymbol{HFI:FIRAS:gain:calibration:accuracy:systematic:217GHz}{}
\setsymbol{HFI:FIRAS:gain:calibration:accuracy:systematic:353GHz}{}
\setsymbol{HFI:FIRAS:gain:calibration:accuracy:systematic:545GHz}{7\%}
\setsymbol{HFI:FIRAS:gain:calibration:accuracy:systematic:857GHz}{7\%}

\setsymbol{HFI:FIRAS:zero:point:accuracy:100GHz:units}{0.8\MJysr}
\setsymbol{HFI:FIRAS:zero:point:accuracy:143GHz:units}{}
\setsymbol{HFI:FIRAS:zero:point:accuracy:217GHz:units}{}
\setsymbol{HFI:FIRAS:zero:point:accuracy:353GHz:units}{1.4\MJysr}
\setsymbol{HFI:FIRAS:zero:point:accuracy:545GHz:units}{2.2\MJysr}
\setsymbol{HFI:FIRAS:zero:point:accuracy:857GHz:units}{1.7\MJysr}

\setsymbol{HFI:FIRAS:zero:point:accuracy:100GHz}{0.8}
\setsymbol{HFI:FIRAS:zero:point:accuracy:143GHz}{}
\setsymbol{HFI:FIRAS:zero:point:accuracy:217GHz}{}
\setsymbol{HFI:FIRAS:zero:point:accuracy:353GHz}{1.4}
\setsymbol{HFI:FIRAS:zero:point:accuracy:545GHz}{2.2}
\setsymbol{HFI:FIRAS:zero:point:accuracy:857GHz}{1.7}

\setsymbol{HFI:unit:conversion:100GHz:units}{0.2415\MJysrmK}
\setsymbol{HFI:unit:conversion:143GHz:units}{0.3694\MJysrmK}
\setsymbol{HFI:unit:conversion:217GHz:units}{0.4811\MJysrmK}
\setsymbol{HFI:unit:conversion:353GHz:units}{0.2883\MJysrmK}
\setsymbol{HFI:unit:conversion:545GHz:units}{0.05826\MJysrmK}
\setsymbol{HFI:unit:conversion:857GHz:units}{0.002238\MJysrmK}

\setsymbol{HFI:unit:conversion:100GHz}{0.2415}
\setsymbol{HFI:unit:conversion:143GHz}{0.3694}
\setsymbol{HFI:unit:conversion:217GHz}{0.4811}
\setsymbol{HFI:unit:conversion:353GHz}{0.2883}
\setsymbol{HFI:unit:conversion:545GHz}{0.05826}
\setsymbol{HFI:unit:conversion:857GHz}{0.002238}

\setsymbol{HFI:colour:correction:alpha=-2:V1.01:100GHz}{0.9893}
\setsymbol{HFI:colour:correction:alpha=-2:V1.01:143GHz}{0.9759}
\setsymbol{HFI:colour:correction:alpha=-2:V1.01:217GHz}{1.0007}
\setsymbol{HFI:colour:correction:alpha=-2:V1.01:353GHz}{1.0028}
\setsymbol{HFI:colour:correction:alpha=-2:V1.01:545GHz}{1.0019}
\setsymbol{HFI:colour:correction:alpha=-2:V1.01:857GHz}{0.9889}

\setsymbol{HFI:colour:correction:alpha=0:V1.01:100GHz}{1.0008}
\setsymbol{HFI:colour:correction:alpha=0:V1.01:143GHz}{1.0148}
\setsymbol{HFI:colour:correction:alpha=0:V1.01:217GHz}{0.9909}
\setsymbol{HFI:colour:correction:alpha=0:V1.01:353GHz}{0.9888}
\setsymbol{HFI:colour:correction:alpha=0:V1.01:545GHz}{0.9878}
\setsymbol{HFI:colour:correction:alpha=0:V1.01:857GHz}{1.0014}